# Ab initio study of phosphorus effect on vacancy-mediated process in nickel alloys – an insight into $Ni_2Cr$ ordering

Jia-Hong Ke[1], George A. Young[1,2], and Julie D. Tucker[1]

[1]Oregon State University, 204 Rogers Hall, Corvallis, OR 97331
[2]Kairos Power, 707 W. Tower Ave., Alameda, CA 94501

## Abstract

The development of long range order in nickel-chromium alloys is of great technological interest but the kinetics and mechanisms of the transformation are poorly understood. The present research utilizes a combined computational and experimental approach to elucidate the mechanism by which phosphorus accelerates the ordering rate of stoichiometric $Ni_2Cr$ in Ni-Cr alloys. A series of Ni-33%Cr-x%P samples (in atomic percent) were fabricated with phosphorus concentrations, x = <0.005-0.1 at.% and aged between 373 and 470°C for times up to 3000 h. The first-principles modeling considers fcc Ni with dilute P as a reasonable approximation for the complex Ni-Cr-P alloy system. Calculation results show a pronounced enhancement of vacancy transport by vacancy-solute pair diffusion via consecutive exchange and rotation jumps of vacancies associated with the phosphorus atom. The energy barriers of these two migration paths are at least 0.35 eV lower than that of vacancy-atom exchange in pure Ni solvent. The analytical diffusion model predicts enhanced solvent diffusion by 2 orders of magnitude for 0.1 at.% P at 400-500°C. The model prediction is in good agreement with the evolution of micro-hardness. We characterize the micro-hardness result by a kinetic ordering model, showing a significant decrease of the activation energy of ordering transformation. These results help gauge the risk of industrial alloys developing long range order which increases strength but degrades ductility and toughness. Specifically, minor alloying additions that bind with excess vacancies and lower the vacancy migration barrier can greatly accelerate hardening via $Ni_2Cr$ precipitation.

## 1. Introduction

Nickel-based alloys are an essential class of structural materials for advanced nuclear power systems and aerospace applications due to their extraordinary combination of creep strength, toughness, and resistance to oxidation and corrosion [1–3]. The superior mechanical properties are mainly attributed to the alloying elements that strengthen materials by the formation of ordered compounds. Nevertheless, it is well known that atomic ordering can lead to increased hardness or embrittlement, which degrades the performance and safe lifetime limits of structural components [4–7]. The ordering transformation mostly occur by the vacancy-atom exchange mechanism, and their kinetics could be affected by vacancy-solute interactions [8,9]. Thus understanding of solute effects on vacancy-mediated processes and predicting atomic ordering development as a function of solute composition are critical considering the complex solution chemistry of most engineering or commercial nickel alloys.

Early studies have showed that small amounts of solutes or impurities (<1 at.%) can modify the rate of alloy phase transformation and property degradation in a significant manner. One noteworthy alloying element demonstrating a pronounced solute effect in nickel-based alloys is phosphorus [10–12], which was found to accelerate the rates of $Ni_2Cr$ ordering and hardening. The acceleration is significant even when the phosphorus composition is dilute. Lehman and Kosel investigated the phosphorus effects on the ordering kinetics of $Ni_2Cr$ alloys under isothermal aging at 500 °C [12]. The ordering rate increased by two times as the phosphorus concentration increases from 0.004 to 0.032 at.% and have a more dramatic increase (~10 times) as phosphorus increased to 0.142 at.%. Likewise, Abd-Elhady and Sargent showed that increasing the phosphorus content in $Ni_2Cr$ alloys (0.002 wt.%, 0.017 wt.%, and 0.078 wt.%) increased the rate of long range ordering at aging temperatures between 425-550 °C [10,11]. For the 0.017 wt.% P alloy the start of the transformation at 450°C was ~2 times faster than the low P alloy, and ~120 times faster for the 0.078 wt.% P alloy. The acceleration is unexpected considering the strong binding between phosphorus and vacancies in both bcc and fcc alloys [13–21] that is generally known to reduce available and freely-migrating vacancies for diffusional processes. The enhanced precipitation or diffusion kinetics via alloying with phosphorus was observed in other alloy systems as well. For example, Kegg *et al.* [22] noted increased precipitation kinetics of $M_{23}C_6$ carbides in an austenitic stainless steel when the alloy was doped with 0.1-0.3 wt.% phosphorus. That research indicated that the accelerated growth rate is likely due to the increase of Cr diffusion in the presence of phosphorus. Additionally, Azarian and Kheloufi reported positron annihilation measurements on a Fe-18Cr-14Ni austenitic steel showing a considerable reduction of vacancy formation energy when alloyed with a dilute amount of phosphorus [23]. The vacancy formation reduced from 1.64 eV in the austenitic steel with <0.005 wt.% P to 1.20 eV in that with 0.05 wt.% P [24]. This energy reduction indicates a strong attractive interaction between phosphorus and vacancy.

The solute-modified ordering kinetics is usually recognized as a result of vacancy-solute interactions that can alter the number of mobile vacancies and kinetic paths to ordering [8,9]. Solutes with strong vacancy binding favors the formation of vacancy-solute pairs but reduce the effective number of vacancies for atomic transport via solute trapping, and therefore the ordering transformation rate decreases [9]. However, the conventional understanding is in contradiction with the accelerated ordering by alloying phosphorus, which is generally known to have strong binding with vacancies in the fcc Ni [17]. Indeed, the previous treatment may not able to produce consistent agreements with experiments over a wide range of solutes and alloys as it neglected other important solute-modified properties, particularly the solute effects on the atom-vacancy exchange frequency and vacancy formation energy in the vicinity of a solute atom. The former involves various types of vacancy jump caused by the addition of solute atoms, while the latter decreases with strong vacancy binding that alters the numbers of vacancy-solute pairs and mobile vacancies. Both effects should play an important role in modifying vacancy-mediated processes as well as kinetic pathways of ordering.

The dramatic modification of hardening and $Ni_2Cr$ ordering rates by a small concentration of alloyed phosphorus poses a serious concern for alloy embrittlement during long-term service and application in nuclear power systems. $Ni_2Cr$ ordering also results in undesirable materials behaviors like lattice contraction, dimensional changes, internal stress, and a change in deformation mode [7,25–28]. Despite these issues, the underlying mechanism for phosphorus-enhanced ordering is not well understood, particularly as the strong solute-vacancy binding is not able to clarify the dramatic acceleration as manifested by experiments [10–12]. Also noted that to the best of our knowledge, both experimental measurement and atomistic modeling of vacancy-mediated diffusion in dilute Ni-P alloys are not available in literature. Given the uncertain role of vacancy binding in vacancy processes and the serious lack of diffusion data of Ni-P dilute alloys, the acceleration of $Ni_2Cr$ ordering by alloying P impurities is still an unresolved question that will be benefited significantly from thorough investigation, taking both vacancy-solute binding and solute-modified vacancy migration into account. It is anticipated that more insights can be gained by applying the approach of analysis to explore other minor solutes of interests. Such information is essential for alloy design consideration and especially valuable for better understanding the role of vacancy-solute interaction in phase transformation kinetics under both aging and irradiation conditions.

In the present study, we investigate the effect of phosphorus by a combination of atomistic simulations, experimental measurement of hardness, and kinetic model of ordering. The fcc Ni with dilute P is treated as an approximation for the complex Ni-Cr-P alloy system in the atomistic simulation, and the phosphorus effect in quasi-random Ni-33 at.% Cr will also be explored. We utilized the five-frequency model of atomic interchange to calculate the diffusion coefficients of Ni, P, and vacancy and their activation energies as a function of phosphorus concentration up to 1 at.%. The energetics required for the substitutional diffusion model were calculated by the first-principles density functional theory (DFT) and climbing image nudged elastic band (CI-NEB) calculations [29,30]. As for the experimental work, isothermal aging of Ni-33 at.%Cr with various content of P from none (<0.005 at.%) to 0.1 at.% was conducted at three temperatures for times up to 3000 h and hardness was measured for all samples. The hardness data was fitted with the kinetic model of ordering by Dienes [31].

The paper is organized as follows: Section 2 details the computational methods, including the atomistic simulations and substitutional diffusion model. This is followed by a description of kinetic models of ordering that are utilized to describe the increase of hardness during aging. Section 3 presents experimental methods, including sample preparation, aging conditions, and hardness measurement. In Section 4 we present the results of first-principles calculations of thermo-kinetic properties, as well as experimental results characterized by kinetic models. In Section 5 we compare our results with previous DFT and experimental studies and discuss the chemical effect of phosphorus on vacancy binding and mobility. Section 5 also presents the similarity of phosphorus effects in fcc Ni and the quasi-random structure of Ni-33 at.% Cr. Finally, we address the limitations and uncertainties. Concluding remarks are given the final section.

## 2. Computational Methods

### *2.1 Diffusion model*

We implemented the multi-frequency framework developed by Lidiard and Le Claire [32–34] to calculate the solute effected diffusion properties mediated by vacancies. For dilute fcc-based solid solutions, the diffusion properties can be predicted by the rates of five distinctive types of vacancy-atom exchanges relative to the solute atom. This particular set of rates is known as the five-frequency model as developed by Lidiard and Le Claire [32,33,35]. The model assumes a dilute solution condition and therefore does not consider solute-solute interactions. Figure 1 illustrates the various diffusion events $\omega_i$ of the five-frequency model following ref. [35]. $\omega_0$ is the hop rate of vacancy-atom exchange in the pure solvent away from any solutes, $\omega_1$ is the vacancy exchange with a 1st nearest neighbor of the solute or vacancy-solute rotation hop, $\omega_2$ is vacancy-solute exchange hop, and $\omega_3/\omega_4$ are the vacancy-solute dissociation/association hops from/to a 2nd, 3rd, or 4th nearest-neighbor solvent atom with respect to the solute atom. Each jump frequency in the five-frequency model can be calculated based on the transition state theory [36,37].

In the Supplementary Information (SI) Section S1, we provide detailed descriptions of the method and equations that are utilized to calculate multiple jump frequencies and phenomenological diffusion coefficients $L_{ij}$. The tracer diffusion coefficient $D_B$ and self-diffusion of the pure solvent atom $D_A(0)$ can be calculated from $L_{ij}$ based on the second-shell approximation [34]. Here in this section we focus on the characterization of solute-modified diffusion properties.

By considering the dilute limit of solute B atoms and pure solvent, $D_B$ and $D_A(0)$ are respectively derived as

$$D_{\mathrm{B}} = 4a_0^{\,2}\omega_2 c_{\mathrm{v}} f_{\mathrm{B}} \exp\left(-E_{\mathrm{vb}}/k_B T\right) \tag{1}$$

$$D_{\mathrm{A}}\left(0\right) = 4a_0^{\,2}\omega_0 c_{\mathrm{v}} f_0 \tag{2}$$

where $a_0$ is the lattice constant, $f_B$ is the correlation factor for solute diffusion as given by $(2\omega_1 + 7\omega_3 F) / (2\omega_1 + 2\omega_2 + 7\omega_3 F)$[32] (see SI Section S1 for the expression of $F$), $f_0$ is the correlation factor of solvent self-diffusion (0.7815 for pure fcc metals), and $E_{vb}$ is the binding energy of the first nearest-neighbor vacancy-solute pair.

Note that if $E_{vb}$ is negative, the attraction or tendency for binding between the solute atom and vacancy favors the formation of vacancy-solute pairs and modifies the number of free vacancies. In the five-frequency model with dilute and isolated solute atoms, the site fractions of bounded vacancies ($c_{\text{pair}}$) and unbounded free vacancies ($c_{\text{v}}$ - $c_{\text{pair}}$) can be determined analytically by the conditions of detailed balance and local thermodynamic equilibrium [34].

$$\frac{c_{\text{pair}}}{12\left(c_{\text{v}}-c_{\text{pair}}\right)\left(c_{B}-c_{\text{pair}}\right)}=\exp\left(-\frac{E_{\text{vb}}}{k_{B}T}\right)=\frac{\omega_{4}}{\omega_{3}} \tag{3}$$

where $c_{\text{B}}$ is the solute concentration and ($c_{\text{B}}$ - $c_{\text{pair}}$) is the unbounded fraction.

The effect of solute on solvent self-diffusion has been well characterized experimentally and theoretically to first order in solute concentration by

$$D_{\text{A}}\left(c_{\text{B}}\right)=D_{\text{A}}\left(0\right)\left(1+bc_{\text{B}}\right) \tag{4}$$

where $D_{\text{A}}(c_{\text{B}})$ is the self-diffusion coefficient of solvent atoms as a function of solute concentration $c_{\text{B}}$, and $b$ is the solvent enhancement factor which can be determined experimentally or theoretically by the multi-frequency of atom jumps near the solute atoms. Several approaches have been taken to calculate $b$ in dilute fcc alloys, including the weak perturbation approximation by Lidiard [38] and calculations built upon the five-frequency model with less restrictive assumptions, such as those by Howard and Manning [39], Ishioka and Koiwa [40], and Bocquet [41]. Here we utilized the self-consistent mean field theory developed more recently by Nastar [42] showing better agreement with Monte Carlo simulations than most of the previous models. Under the approximation restricted to nearest-neighbor effective interactions, the model offers an analytical form of the enhancement factor $b$ as a function of the five frequencies [42]. The analytical expression of $b$ factor, including the contributions of frequency enhancement and correlation enhancement, is provided in SI Section S2.

In additional to the solvent enhancement factor, the role of solute in vacancy-mediated processes can be manifested by the effective vacancy diffusion coefficient $D_{\text{v}}^{\text{eff}}$, which includes both the contributions of unbounded free vacancies and vacancy-solute pairs. This effective coefficient is defined by the vacancy flux in the absence of solute concentration gradient, $J_{\text{v}}=-D_{\text{v}}^{\text{eff}}\nabla c_{\text{v}}$, where $D_{\text{v}}^{\text{eff}}$ is obtained as [34]

$$D_{\text{v}}^{\text{eff}}=\frac{k_{B}T}{n}\frac{c_{\text{B}}L_{\text{vv}}-c_{\text{pair}}L_{\text{vB}}}{c_{\text{v}}c_{\text{B}}-c_{\text{pair}}{}^{2}} \tag{5}$$

$L_{vv}$ and $L_{vB}$ are phenomenological coefficients which are related to $L_{AA}$, $L_{BB}$, and $L_{AB}$ by $L_{vv} = L_{AA} + L_{BB} + 2L_{AB}$ and $L_{vB} = -(L_{BB} + L_{AB})$, respectively. Note that the diffusion coefficients of free vacancies ( $D_{\mathrm{v}}^{\mathrm{free}}$ ) and vacancy-solute pairs ( $D_{\mathrm{s\text{-}v}}^{\mathrm{pairs}}$ ) can be derived respectively as

$$D_{\mathrm{v}}^{\mathrm{free}} = \frac{k_B T L_{AA}^{f}}{n c_{\mathrm{v}}'} \tag{6}$$

$$D_{\mathrm{s\text{-}v}}^{\mathrm{pairs}} = \frac{k_B T L_{BB}}{n c_{\mathrm{pair}}} \tag{7}$$

where $L_{AA}^{f}$ is the free vacancy contribution of the $L_{AA}$ (see SI Section S1), which gives

$2ns^2 c_{\mathrm{v}}' \omega_0 / k_B T$ .

*2.2 First-principles methodologies*

The vacancy-phosphorus interaction and the migration barriers described in Section 2.1 are calculated using the Vienna *Ab initio* Simulation Package (VASP) [43,44], a quantum mechanical density functional theory-based code. All calculations were spin-polarized and initialized under the ferromagnetic ordering. We adopted the previous approach of *ab initio* calculations predicting diffusion properties of dilute Ni-based alloys [45,46]. The plane wave energy cutoff was selected to be 479 eV for all calculations. The interactions between ions and core electrons are described using the projector augmented wave (PAW) method [47,48] and the Perdew-Burke-Ernzerhof [49] parameterization of the generalized gradient approximation (GGA) was used for the exchange correlation potentials. The k-point mesh is generated automatically using the Monkhorst-Pack scheme. The k-point mesh describes the sampling of the Bravais lattice and a 4×4×4 mesh was used for all calculations. Calculations were performed with a 108 atom periodic simulation cell. The simulation cell is a 3×3×3 supercell of the fcc conventional cell with a lattice parameter of 3.52 Å for pure Ni. The convergence tests for these values can be found in previous work [45,46]. In all CI-NEB runs, the break condition of energy convergence for electronic relaxation is $10^{-5}$ eV and that of force for ionic relaxation is 0.01 eV/Å.

The vacancy-solute binding energy is calculated by the following equation [50],

$$E_{\mathrm{vb}}^{\mathrm{P}} = E\left(106\mathrm{Ni}{+}1\mathrm{P}{+}1\mathrm{v}\right)_{\Omega} - E\left(107\mathrm{Ni}{+}1\mathrm{P}\right)_{\Omega'} - E\left(107\mathrm{Ni}{+}1\mathrm{v}\right)_{\Omega} + E\left(108\mathrm{Ni}\right)_{\Omega'} \qquad (8)$$

where P denotes the solute species (phosphorus) and v denotes the vacancy. The energy defined in the form of Eq. (8) is referred to as the indirect binding energy [51]. The energy of the isolated vacancy and solute atom are subtracted separately from the energy of the system where they interact, and then the energy of a perfect, un-defected crystal is added for mass balance. Note that we adopt the sign convention different from ref.[50]. The negative binding energy in this work indicates the energetically favorable binding of a solute with a vacancy. Calculations of the indirect binding energy were performed with the fcc supercell where the volume and shape are fixed but ionic relaxations are allowed. Note that the volumes of the solute-containing simulations cells are identical with the volumes of the pure Ni cells, regardless of the presence of the vacancy (Ω and Ω' for the volume with and without a vacancy, respectively). Bader charge analysis is performed with the electron charge density converged until the energy is below $10^{-6}$ eV. The tool developed by Henkelman *et al.* [52,53] with the algorithm proposed by Yu and Trinkle [54] for charge integration is implemented.

Migration barriers for vacancy hopping between the two lattice sites were calculated by using the CI-NEB method [29,55] with three intermediate migration images for solvent self-diffusion and solute-vacancy exchange hop, and five images for other jumps of solvent atoms and the vacancy near the solute atom. A cubic spline was fitted to the migration energy landscape and migration barriers were determined by taking the energy difference between the saddle point with respect to the energy at the first or final lattice point. The saddle point represents the configuration where the energy is maximum in the migration energy profile.

*2.3 Kinetic models of ordering*

Two kinetic models were utilized in this study to describe the kinetics of ordering transformation, including the stoichiometric theory of Dienes based on an analogy of the chemical reaction rate [31] and a modified Kolmogorov–Johnson–Mehl–Avrami (KJMA) model [56]. The stoichiometric theory was later re-derived by Vineyard for the direct interchange and vacancy mechanisms [57]. Note that due to the similar X-ray scattering properties of Ni and Cr, characterization of ordering transformation of $Ni_2Cr$ using X-ray diffraction analysis is challenging [58]. Thus indirect methods measuring the change of a materials property, such as resistivity, lattice constant, and hardness were employed mostly in previous studies [10,12,25,27,59,60]. Here we use the experimental data of measured hardness to quantify the evolution of ordering process and fit the data by the kinetic models and associated parameters. Experimental methods are provided in the next section.

Similar to the previous treatment quantifying ordering by the property change, we assume that the fraction transformed is equivalent to the change in hardness and the modified KJMA equation of ordering is given as:

$$\eta = \frac{H - H_0}{H_{\max} - H_0} = 1 - \exp\left(-kt^n\right) \tag{9}$$

where $\eta$ is a parameter characterizing the ordering, $n$ is the Avrami exponent, $H$ is the measured hardness, $H_{max}$ is the maximum and saturated value, and $H_0$ is the initial value. $H_{max}$ and $H_0$ are determined based on the hardness data for each Ni-P alloy with the same P composition. $k$ is a kinetic coefficient described by an Arrhenius form as $k = k_0 \exp\left(-Q/k_B T\right)$, where $Q$ is the activation energy and $k_0$ is the pre-exponential factor.

Based on the kinetic theory of Dienes [31], the ordering rate is given as

$$\frac{d\eta}{dt} = v \exp\left(-U/k_B T\right)\left\{\xi\left(1-\xi\right)\left(1-\eta\right)^2 - \left[\eta + \xi\left(1-\xi\right)\left(1-\eta\right)^2\right]\exp\left(-W/k_B T\right)\right\} \tag{10}$$

where $v$ is the frequency of vibration associated with the interchange between an atom and a vacancy, $U$ is the activation energy, $\xi$ is the average concentration at stoichiometry, which is 1/3 for $Ni_2Cr$ ordering. $W$ is the ordering energy defined as an expansion of $\eta$ [31,61]:

$$W = W_0\left(\eta + \eta^3/3\right) \tag{11}$$

As noted by Vineyard [57] for ordering via the vacancy-atom interchange mechanism, the activation energy $U$ should include the contribution of vacancy formation plus the activation energy of vacancy migration. The numerical solution of Eq.(10) was solved by MATLAB software implementing the Runge-Kutta method with a variable time step.

Note that Dienes' theory and the modified KJMA model have both been successfully applied to study the ordering kinetics of $Ni_2Cr$, including the previous studies by Barnard *et al.*[62], Young *et al.*[6,7], and Makarova and Paskal [61]. In this work we considered the parameters used in those studies as reference values and treated the activation energy of each model as a fitting parameter. The fitting parameters were calibrated by the experimental measurement of hardness as a function of aging time for each dilute composition of alloying phosphorus. The results using the stoichiometric theory are shown in Section 4.3 and the modified KJMA modeling results are summarized in SI Section S4.

## 3. Experimental Methods

The effect of phosphorus was investigated by fabricating and aging a series of alloys near the $Ni_2Cr$ (atomic) stoichiometry with ~33 at.% Cr. The alloys were fabricated with high purity starting materials such that the $Ni_2Cr$ binary alloy contained <0.005 wt.% phosphorus and five conditions were fabricated, starting with the $Ni_2Cr$ (atomic) composition: no addition of phosphorus, 0.01 at.% P, 0.05 at.% P, 0.075 at.% P, and 0.1 at.% P. The alloys were arc melted, homogenized at 1090 °C for 24 h, and hot rolled at ~1000 °C in three passes from ~10 mm to ~5 mm thick plate. Finally, the plate was annealed at 1090 °C for 1 h and water quenched. From each plate, rectangular coupons were saw cut for isothermal aging. Isothermal agings were conducted at 373 °C, 418 °C, and 470 °C for times up to 3000 h. After aging, each sample was mounted, ground, and polished to a metallographic finish to facilitate micro-hardness testing. The micro-hardness tests were conducted with a Vickers indenter using 500 grams-force. The samples were metallographically finished to 0.3 microns, and the Vickers micro hardness dwell time was the machine nominal setting of 15 seconds. For each sample the average of the five readings were used to determine the average hardness of the sample. The typical standard deviation in hardness was ±7 HV, with a maximum standard deviation in all the samples being ±18.1 HV. Thus, the change in hardness from the water quenched unaged condition was used to assess the evolution of long range order.

## 4. Results

### *4.1 Thermo-kinetic parameters of dilute Ni-P calculated by first principles*

The thermo-kinetic parameters and data obtained by the first-principles DFT calculations are summarized in Table 1. The vacancy-phosphorus binding energy is determined by the two methods: the ratio of dissociation/association frequency following Eq. (3) and the indirect method via Eq. (8). Both methods show similar vacancy binding energy of −0.29 - −0.30 eV with an energy difference less than 0.015 eV. In this paper, we use the sign convention by which negative binding energy indicates the tendency of attraction or binding of a solute with a vacancy. The calculation result shows that P-vacancy binding is energetically more favorable compared to other common solutes including Si, Cr, Fe, Mn, and Al. These solutes demonstrate either relatively weaker vacancy-solute interaction or even slight repulsion [63]. One element showing stronger vacancy-solute binding is S (sulfur) with a binding energy of −0.42 eV [64]. Note that the 3*sp*-nonmetal impurities (Si, P, and S) in fcc Ni follows the relative order of vacancy binding magnitude: $\left|E_{\mathrm{vb}}^{\mathrm{Si}}\right| < \left|E_{\mathrm{vb}}^{\mathrm{P}}\right| < \left|E_{\mathrm{vb}}^{\mathrm{S}}\right|$ , which is the same ordering as manifested in the first-principles studies on bcc Fe [13,14,18,21].

The migration barriers of the five distinctive types of vacancy-atom exchanges are listed in Table 1, and Figure 2 shows their minimum energy paths obtained by the CI-NEB method. The calculation result of $\omega_0$ migration barrier in pure Ni is consistent with experimental measurements [65] and both the barrier energy and attempt frequency of Ni diffusion agree well with our previous study, which were determined as 1.09 eV and $4.48 \times 10^{12}$ Hz, respectively [45]. It is worth noting that both $\omega_1$ and $\omega_2$ migration barriers are lower than $\omega_0$ by more than 0.35 eV. This result indicates that a vacancy possesses much higher hopping frequencies when rotating around and exchanging with the P solute atom, which are respectively manifested by the low $\omega_1$ and $\omega_2$ migration barriers. The higher $\omega_3$ migration barrier than $\omega_4$ for all vacancy-solute dissociation/association hops shows that it has a higher probability to form bounded vacancy-solute pairs than that generated by random process. Considering the strong tendency of vacancy-phosphorus binding, the high $\omega_1$ and $\omega_2$ frequencies suggest a fast solute diffusion mechanism operated by solute-vacancy pairs, which is also known as the Johnson mechanism [66]: diffusion proceeds by cycles of exchange (inversion) and rotation (re-orientation) of the solute-vacancy pair. It is therefore expected that the activation energy of phosphorus diffusion should be much lower than that of solvent diffusion. Note that the effective diffusion via solute-vacancy pairs, which is associated with the transport of vacancies, contributes further to the enhanced mobility of vacancies and solvent atoms. The overall calculated thermo-kinetic parameters reveal that the vacancy-solute pair diffusion may be involved in the accelerated ordering transformation in Ni alloys by alloying with phosphorus.

The quantitative analysis of the solute diffusion and phosphorus effect on vacancy-mediated diffusion will be presented in the next section with the consideration of the correlation effect.

*4.2 Vacancy-mediated diffusion*

Following the five-frequency model, the diffusion coefficient of P in the dilute Ni-P alloy can be obtained. Note that in this study, the pure Ni vacancy formation energy $G_{\mathrm{vf}}$ was adopted from the previous work as determined by utilizing the $H_{\mathrm{vf}}$ value from experiments and fitting $S_{\mathrm{vf}}$ to reproduce the experimental diffusion data, which respectively gave 1.79 eV and 5.71 $k_B$ [45]. The vacancy concentration in pure solvent under thermodynamic equilibrium can then be determined by $\exp(-G_{vf}/k_BT)$. Figure 3 shows the temperature dependence of the P solute tracer diffusion coefficient in Ni and self-diffusion coefficient of pure Ni. The diffusion coefficient of P solute is significantly higher than that of self-diffusion of pure Ni. The difference is expected considering the lower $\omega_1$ and $\omega_2$ migration barriers than $\omega_0$ and the tendency of vacancy-solute binding. The calculated activation energy of Ni self-diffusion is 2.85 eV (the same as $H_{vf} + E_0$), which is close to the recommended value 2.88 eV [65] and agrees well with most of the experimental studies showing that the range is between 2.7-3.0 eV [67]. The calculated activation energy of P tracer diffusion in dilute Ni-P is 2.16 eV, which is ~0.7 eV lower than Ni self-diffusion. As shown in the previous section, it is expected that the low migration barriers of both $\omega_1$ (rotation) and $\omega_2$ (exchange) hops and vacancy-solute binding contribute to the fast solute diffusion by vacancy-solute pairs. The associated activation energy can then be approximated as $H_{vf} + E_{vb} + \max(E_0, E_1) \approx 2.20$ eV, which is close to the calculation result, 2.16 eV. The ~0.04 eV error is likely due to the rough approximation of the third term ($\max(E_0, E_1)$), whose accurate estimation should involve the correlation effect of solute diffusion ($f_B$). Note that the fast diffusion of P in fcc Ni is similar to that of S, which is known as a fast diffuser in fcc Ni according the experimental data [68] and a recent DFT study [64]. The study reports remarkably low $\omega_1$ and $\omega_2$ migration barriers and strong vacancy binding calculated by first-principles DFT [64].

Figure 4(a) shows the temperature dependence of solvent Ni diffusion coefficient in Ni-P alloys with various P solute contents up to 1 at.%. The result is calculated by using the solvent enhancement factor *b* as described in SI Section S2. A significant enhancement effect of P solute on Ni self-diffusion is demonstrated even as P composition is less than 0.1 or 0.01 at.%. The enhancement can reach by a factor of 10 and $10^2$ when the composition of P is 0.01 and 0.1 at.% at 500 °C, respectively. Note that the effect of solute enhancement or *b* factor is more pronounced at low temperature due to the strong tendency of vacancy-phosphorus binding, which promotes efficient diffusion via vacancy-solute pairs. The trend can also be seen in Figure 4(b) showing the enhancement of Ni diffusion as a function of P composition at temperatures from 300 to 700 °C. Note that phosphorus is typically recognized as a detrimental impurity in nickel-based superalloys and hence its nominal composition is generally specified at less than 0.015 wt.% in industrial applications. It is indicated in Figure 4(b) that at this specified composition, the Ni diffusion enhancement is minor when the temperature is higher than 700 °C. The enhancement is less than 2 in Ni-0.01 at.% P at 700 °C, and thus the P effect on diffusion properties can be neglected for most of the high-temperature applications. However, such an effect will be significant if the P composition is not well restricted below 0.01 at.% for long-term and low-temperature applications like those in nuclear power systems, where the typical operating temperature is 300-400°C.

To explore the microscopic mechanism of solvent enhancement and characterize the vacancy mobility, Figure 5(a) shows the temperature dependence of the effective vacancy diffusion coefficient $D_{v}^{eff}$ (Eq. (5)) in Ni-P alloys with various P solute contents. The plots for the diffusion coefficients of free vacancies ( $D_{v}^{free}$ , Eq.(6)) and vacancy-solute pairs ( $D_{s\text{-}v}^{pairs}$ , Eq.(7)) are shown as symbols in Figure 5(a). The plots of $D_{v}^{free}$ and $D_{s\text{-}v}^{pairs}$ are manifested as the two limits of vacancy transport: one via freely-migrated vacancies and the other via bounded vacancy-solute pairs. Increasing the P composition from pure Ni to dilute Ni-P alloys clearly increases the effective vacancy mobility. Additionally, as P composition increases, $D_{v}^{eff}$ (solid lines in Figure 5(a)) increases in magnitude from the limit of freely-migrated vacancies and gradually approaches to the limit of vacancy-solute pairs. The result shows that diffusion by vacancy-solute pairs becomes the dominant vacancy transport mechanism as the P composition increases to 1 at.%. Note that the $D_{v}^{eff}$ plots in Figure 5(a) are non-Arrhenius at a certain range of temperature. The non-Arrhenius feature is attributed to the change of the dominant vacancy diffusion mechanism as temperature varies, as clearly seen in Figure 5(b). Increasing temperature increases the probability of kinetic events for atoms to overcome the barrier of vacancy-solute dissociation hops, thereby reducing the bounded number of vacancy-solute pairs and the contribution of pair diffusion relative to that by free vacancies. It is noted that for high phosphorus Ni alloys like Ni-1 at.%P, vacancy diffusion by vacancy-solute pairs contributes to more than 90% for temperature lower than 700 °C. Removal and control of phosphorus composition to the order of 0.001 at.% is required to ensure a reduced impact of such a undesirable effect for low-temperature applications.

*4.3 Hardness data and kinetic modeling*

The micro-hardness change of the five experimental alloys during isothermal aging is summarized in Table 2 (also see SI Section S3 for the actual micro-hardness). As shown, alloyed phosphorus has a marked effect on the rate of hardening via long range order, where the highest phosphorus alloy exhibits ~3 orders of magnitude faster hardening kinetics than the alloy with no intentional addition of phosphorus. Even at a phosphorus composition as low as 0.01 at.% (0.006 wt.%), the hardness increases much more rapidly than that of Ni-33at.%-Cr without alloying phosphorus. For example, for Ni-33at.%Cr aging at 373 °C for 1000 h the change of hardness is 10.4 HV, while for the alloy with 0.01 at.% phosphorus the change of hardness is 58.4 HV. To demonstrate the acceleration effect by phosphorus in a clear manner, Figure 6 plots the effect of alloyed phosphorus (in at.%) on the time to change the as-received hardness by +50 HV during isothermal aging at 470 °C. The order of acceleration factor can be well approximated and fitted linearly as a function of alloyed phosphorus concentration. As shown, the acceleration due to phosphorus from ~0 to the maximum impurity limit (~0.015 wt.% or 0.027 at.%) in industry Ni alloys would be on the magnitude of 3.6. While the acceleration is not as significant as the higher phosphorus alloys, it still could result in heat-to-heat variability in ordering kinetics. In addition to the acceleration effect, it is noted that the peak hardness after aging at each temperature increase slightly with the alloyed phosphorus even at a dilute content like 0.01 at.%. This trend is likely attributed to the larger fraction of ordered domain by alloying phosphorus due to the faster ordering transformation.

Figure 7 (a)-(c) shows the fraction of measured hardness increase as a function of aging time at 373, 418, and 470°C, respectively, which is used to characterize the evolution of ordering transformation. The result demonstrates pronounced acceleration of hardening caused by the addition of phosphorus, and the acceleration increases with P composition. The hardness data at 470°C shows that the hardening rate of Ni-33 at.% Cr alloys with 0.1 at.% P exhibits ~2 orders of magnitude faster than that without phosphorus. The increased rate is consistent with the experimental results by Abd-Elhady and Sargent [10,11] showing that the ordering transformation at 450°C is ~120 times faster for the 0.078 wt.% P alloy (~0.14 at.% P). Note that the data reported by Abd-Elhady and Sargent indicates stronger acceleration by alloying phosphorus at 450°C than at 500°C. Likewise, the stronger acceleration at lower temperature is also observed in our hardening data, showing that at 373°C the acceleration of hardening increase by phosphorus is more pronounced compared to that at 473°C (Figure 7(c)). The hardening rate of Ni-33%Cr alloys with 0.1% P at 373°C exhibits ~3 orders of magnitude faster than that without phosphorus, as clearly seen in Figure 7(a).

The increase of ordering acceleration as temperature decreases suggests the change of kinetic mechanism by the modification of activation energy. Although the addition of alloying solute or impurity atoms is able to modify both kinetic and thermodynamic properties, the latter influence is expected to be small in the present study. As the phosphorus contents in the present and previous experiments are in dilute amounts (≤ 0.1 at.%), the phosphorus effect on the thermodynamic quantity, such as the ordering energy and disorder-order transition temperature, can be considered marginal, and hence the effect on thermodynamics quantities of ordering can reasonably be neglected.

By using the Dienes kinetic model of ordering (Section 2.3) and treating the activation energy $U$ in Eq.(10) as an adjusting parameter to fit with the hardening data, the change of $U$ as a function of the phosphorus composition can be estimated. Note that in the present study, it is assumed that the phosphorus effect on ordering is by modifying the diffusion mechanism in the disordered matrix, and thus the effect on $U$ is approximated as linearly dependent of disordered fraction ($1-\eta$). Dienes theory shows that initial fluctuation of ordering is required to have a nonzero ordering rate when $\eta = 0$ [31], as can be seen in Eq.(10). A small initial fluctuation $10^{-3}$ is used in the calculation. The increase of fluctuation reduces the time of onset of nucleation, but does not affect the latter trend of ordering. It was found that $10^{-3}$ is able to produce consistent agreement in ordering evolution for all cases in the study. Note that $10^{-3}$ is about an order of magnitude smaller than that used by Dienes [31]; however, we found that a large fluctuation is not required here possibly due to the >100°C undercooling (373-470°C) from the order-disorder transition temperature (580-630 °C) of $Ni_2Cr$.

The solid lines in Figure 7 (a)-(c) show the simulation results and Table 3 lists the parameters used in the model. We adopted the parameters derived by Makarova and Paskal [61] except the activation energy $U$ and initial ordering fluctuation. We find the optimal fitting values for activation energy $U$ to obtain ordering consistent with experiments, as summarized in Table 3 for each Ni-P alloy. For 0% P the fitted activation energy $U$ is 2.26 eV, which is 5.6% larger than that obtained by Makarova and Paskal [61]. The discrepancy is acceptable considering the uncertainties caused by alloy fabrications. Increasing phosphorus composition from 0% to 0.1% causes a significant reduction (~0.7 eV) of activation energy $U$ from 2.26 eV to 1.57 eV. Note that the similar results can be obtained by utilizing the KJMA equations for data fitting. The result is shown in SI Section S4.

## 5. Discussion

### *5.1 Insights of the first-principle-based modeling and comparison to experiment*

The first-principle calculations in this study shows that a minor amount of phosphorus is able to increase the mobility of vacancies in fcc Ni significantly via the diffusion of vacancy-solute pairs or Johnson mechanism [66]. The model prediction is consistent with the accelerated ordering transformation by phosphorus, as observed in previous [10–12] and this studies. The enhancement in vacancy mobility and solvent diffusion by phosphorus is attributed to the strong vacancy binding as well as lower $\omega_1$ and $\omega_2$ barriers. Increasing the phosphorus composition from 0 to 0.1 at.% reduces the activation energy of solvent diffusion from that of pure Ni self-diffusion (2.88 eV) to phosphorus solute diffusion (~2.20 eV), as can be seen in Figure 3 and Figure 4(a). It appears that this 0.68 eV reduction is in good agreement with the modeling results of kinetic ordering, showing that the best fit to our hardness data from 0 to 0.1 at.% P requires a ~0.7 eV reduction of activation energy, as described in Section 4.3. We note that this excellent agreement in quantity is likely fortuitous due to the approximate treatment of the ordering theory based on chemical reaction rates, which for example, does not treat the variation of activation energy associated with the complex Ni-Cr-P-vacancy configurations. Specifically, Ni-33 at.% Cr is a highly concentrated Ni-Cr alloy and the change of migration barriers associated with phosphorus atoms should be affected by the evolution of local chemical environment. Nevertheless, analytical expressions of diffusion models characterizing the effect of local environment are not available for concentrated alloys. A more rigorous approach like cluster expansion could be utilized to describe the local environment dependency by an effective Hamiltonian description of activation energies [69,70], but such a study in beyond the scope of this work. In Section 5.4, the phosphorus effect in concentrated Ni-33 at.% Cr will be discussed by the approach of DFT calculations for the quasi-random structure of Ni-Cr-P with a vacancy.

Despite the limitations mentioned above, the first-principles calculation clearly suggests a pronounced phosphorus-enhanced solvent diffusion in fcc Ni, and it is expected that the qualitative conclusions derived here can apply to most Ni-based fcc alloys. One important insight given by this study is on the role of vacancy-solute binding in long-range ordering. Solutes or impurities with strong vacancy binding have long been known to trap vacancies, thereby modifying ordering pathways and reducing ordering kinetics. Note that several experimental work indicated that the short-range ordering (SRO) kinetics of quenched Cu-based alloys can be influenced by vacancy-solute complexes through the modification of defect annealing or decay kinetics [71–73]. Their results suggested simultaneous contributions of vacancy-solute pairs and free vacancies to the observed SRO kinetics. In the present study we make a clarification that vacancy-solute binding is able to accelerate ordering kinetics as long as the vacancy-solute pairs are mobile. The mobility of vacancy-solute complexes is determined by the ability to keep vacancy in the nearest neighbor of solutes as well as the migration barriers of vacancy-solute exchange and vacancy rotation around the solute atom. The bounded solute pairs can be considered as immobile traps if either one of the exchange ($\omega_2$) or rotation ($\omega_1$) barrier is higher than the pure solvent atoms ($\omega_0$).

Practically for industrial alloys like Alloy 690 and its weld metals, the observation of increased ordering kinetics with alloyed phosphorus at the levels investigated indicates that evident heat-to-heat variability in ordering kinetics could be observed, as indicated in Figure 6 and Figure 7. Furthermore, similar solutes such as sulfur could have similar effects. The composition dependence of the ordering rate could be significant and larger effects could be observed in fusion welds, where phosphorus and sulfur may strongly segregate to inter-dendritic regions, locally increasing the concentration.

Note that these thermo-kinetic parameters represent valuable information not only for the dilute solute diffusion database but also alloy design and reliability evaluation for long-term applications. For example, it has been shown experimentally by Garner and co-workers [74–78] that the addition of certain fast diffusers or solutes, including phosphorus at a very low level, is able to suppress radiation-induced void swelling in Fe-Ni-based austenitic steels. Although without modern atomistic calculations of migration barriers at the time of experiments, the authors suggested the potential influence of phosphorus on the enhanced vacancy diffusion, which potentially reduces vacancy supersaturation during irradiation. The enhancement of vacancy-mediated diffusion in Fe-Ni austenitic steels by the addition of phosphorus was also observed by Dean and Goldstein [79,80] in the direct measurement of interdiffusion coefficients at temperatures between 610 and 875 °C, and the diffusion enhancement was found to be stronger at lower temperature and higher phosphorus level. It is thus suggested the qualitative similarity of phosphorus-enhanced vacancy-mediated diffusion in Ni-based and Fe-based fcc alloys. Nevertheless, austenitic steels possess a very different magnetic state (paramagnetic) under thermodynamic equilibrium compared to fcc-Ni alloys, giving rise to the complexity caused by the local magnetic moment. Thus more systematic assessments and calculations would be valuable to offer a comprehensive view on the solute effects in different fcc alloys.

*5.2 Vacancy-phosphorus interaction*

It has been shown in earlier studies that solute or impurity atoms having a vacancy binding energy more negative than −0.2 eV will induce the formation of vacancy clusters under non-equilibrium conditions such as cold work or irradiation [81,82]. The first-principles calculation of vacancy-phosphorus interaction in this study is in generally good agreement with the previous experimental work, showing the strong tendency of phosphorus atoms to interact and form vacancy aggregation under electron irradiation [17]. Similar results suggesting the strong vacancy-phosphorus interaction were also found in austenitic steels [16,23,24].

To distinguish the origin of the vacancy-phosphorus binding, the total binding energy is decomposed into distortion and electronic binding energy. The distortion binding energy is defined as the energy contribution associated with the lattice relaxation when forming a vacancy-solute pair without taking chemical interaction between the solute and solvent atoms, while the electronic binding energy is the rest of the contribution [13,83]. Based on the calculation from Ref. [13,83], the distortion binding energy of the vacancy-phosphorus pair is calculated as −0.024 eV, which contributes to only ~8% of the total binding energy. The result shows that the electronic or chemical effect plays a more dominant role.

The electron or chemical effect can be visualized more clearly in Figure 8 (a) and (b) showing the comparison of the charge density contour maps in the (111) plane calculated with and without including the vacancy, respectively. The Ni-P bond in the nearest distance with the vacancy in Figure 8 (b) possesses a higher charge density than the Ni-P bond in Figure 8 (a) at the bond critical point, which is defined as the saddle point of the charge density with one positive and two negative curvatures [84]. This characteristic suggests a higher strength of the shared or covalent interaction of Ni-P when bounded with a vacancy based on the quantum theory of atoms in molecules [84]. The stronger Ni-P bond when paired with a vacancy also supports the attractive vacancy-phosphorus binding. The results of Bader charge analysis are displayed in Figure 8 (a) and (b). The Bader charge of the phosphorus atom is about −0.3*e* in both cases, which appears to be affected weakly by the vacancy. Note that the negative Bader charge of P indicates that the phosphorus atom attracts valence electrons from the Ni solvent atoms. This result is similar to sulfur having a negative Bader charge −0.4*e* but in opposite to Al which has a positive charge of +1.9*e* [64].

*5.3 Phosphorus effect on vacancy migration barriers*

As discussed above in Section 4.1 and 4.2, the phosphorus-enhanced vacancy mobility and solvent diffusion are attributed to the lower migration barriers of both rotation ($\omega_1$) and exchange ($\omega_2$) hops than that of the vacancy exchange in pure solvent ($\omega_0$), which together enable the effective operation of vacancy-solute pair diffusion. These characteristics are similar to the sulfur effect in fcc Ni, as noted by a first-principles study [64] showing exceptionally low $\omega_1$ and $\omega_2$ migration barriers and large solvent enhancement factor *b*. In this section we further clarify the similarity of their influence on migration barrier by probing the electronic origin and comparing that with Lomaev *et al*. [64].

The charge density contour presented in the recent first-principles study [64] demonstrated that the ultrafast sulfur diffusion is caused by the enhancement of Ni-S chemical bonding at the saddle point, which reduces the $\omega_2$ migration barrier, whereas for aluminum diffusion the enhancement of bonding at the saddle point is absent, and thus the $\omega_2$ migration barrier is higher than sulfur. The enhancement of Ni-S bonding was manifested by the strong sharing of electrons between the migrating atom and the nearest atoms at the saddle point. This covalent characteristic occurs at the transition state of Ni-P diffusion as well, as clearly seen in Figure 9 (a) and (b) for the electron charge density contour map of rotation ($\omega_1$) and exchange ($\omega_2$) hops at the saddle point, respectively. Note that the charge density contour shown in Figure 8 (b) is symmetrically equivalent to the initial state of $\omega_1$ and $\omega_2$ hops. The electronic structure feature suggests the similarity of phosphorus and sulfur for their influence on vacancy mobility and vacancy-solute binding.

Although both P and S show a similar trend of significantly reduced $\omega_1$ and $\omega_2$ migration barriers compared to $\omega_0$, one discrepancy is that the $\omega_1$ and $\omega_2$ migration barriers for Ni-S (reported in ref. [64]) are further evidently lower than Ni-P (this study). The $\omega_1$ and $\omega_2$ migration barriers were calculated as 0.45 and 0.25 eV, respectively [64]. The discrepancy in energy magnitude may be caused by the difference in electron transfer and bonding/hybridization characteristics between the saddle point and initial state. The Bader charge of P at the transition state of the $\omega_2$ hop is −0.05*e* (shown in Figure 9 (b)), which in magnitude is significantly lower than that during the initial state (−0.29*e* shown in Figure 8 (b)). The reduction is possibly caused by the change of the first-neighbor coordinate number decreasing from 11 at the initial state to 4 at the transition state. However, the role of electron transfer in the migration barrier remains unclear.

Also note that according to an early first-principles study of transition metal solute diffusion in Ni by Janotti *et al.*[84], the presence of directional bonding involving the *d* electrons of solutes can reduce the compressibility of solutes, thereby increasing the migration barrier of solute diffusion. The effect is different from the bonding of *sp* solute which lowers the migration barrier, as noted by Lomaev *et al.* [64]. It was suggested by the same authors that the electronic effect plays a more dominant role in diffusion of either *sp* elements in transition metal matrix or transition metal elements in *sp*-matrix, whereas the diffusion of *sp*/transition element in *sp*/transition metal matrix is controlled by the size effect [64,85]. Additional exploration of chemical bond analysis would be valuable to unravel the similarities and disparities among various groups of solute elements, but further discussion of this topic is beyond the scope of this work.

*5.4 Phosphorus effect in concentrated Ni-33Cr alloys*

Despite that phosphorus-accelerated $Ni_2Cr$ ordering was observed in both current and previous experimental studies, the phosphorus effect on the microscopic vacancy mechanism in concentrated Ni-33 at.% Cr alloys remains uncertain due to the chemical and configurational complexity. Specifically, the role of dilute phosphorus in vacancy-solute pair diffusion, which is shown as an efficient diffusion pathway in fcc Ni (see Section 4.2), has not been confirmed in the concentrated Ni-Cr. To verify that vacancy-phosphorus pair diffusion can accelerate vacancy-mediated processes in Ni-33 at.% Cr in a manner similar to fcc Ni, we perform additional DFT calculations for the quasi-random structure of Ni-Cr-P-Vacancy system.

The special quasi-random structure (SQS) of the 2×2×2 fcc supercell is generated by the Monte Carlo simulation code (*mcsqs*) built within the Alloy Theoretic Automated Toolkit (ATAT) [86], seeking the best match of correlation functions to a target random alloy considering the first two nearest-neighbor shells of pairs and triplets. The generated SQS consists of 20 Ni atoms, 10 Cr atoms, 1 P atom, and 1 vacant site (Vac), as displayed in Figure 10(a). Note that to assess the vacancy-solute pair diffusion in the concentrated Ni-Cr alloys, the structure containing a first nearest-neighbor vacancy-phosphorus pair was selected artificially during the Monte Carlo searching of SQS. By doing so the 20Ni-10Cr-1P-1Vac SQS can be used to calculate vacancy-phosphorus binding energy and migration barriers of rotation ($\omega_1$-type) and exchange ($\omega_2$-type) hops in the vicinity of the P atom, which are the three key factors determining the contribution of diffusion by vacancy-solute pairs. Note that for the 20Ni-10Cr-1P-1Vac quasi-random alloy, there are four distinctive rotation hops around the P atom, as shown by the four highlighted atoms (including 3 Ni atoms and 1 Cr atom) as shown in Figure 10(a). These four atoms can exchange with the vacant site while keep nearest-neighbor with the P atom. The phosphorus effect on vacancy kinetics can then be manifested by comparing the associated migration barriers in 20Ni-10Cr-1P-1Vac SQS to the same structure but with the P atom replaced by Ni or Cr.

The parameters of DFT and CI-NEB calculations as described in Section 2.2 were used, including the energy cutoff, k-points per reciprocal atom, initial magnetic moments, PAW potentials, and condition of energy convergence. For the spin-polarized DFT calculations of concentrated Ni-Cr alloys, we adopted the treatment from ref. [87] by assuming that the initial magnetic moments of Ni and Cr atoms are respectively +1 and −1 $\mu_B$. CI-NEB calculations for SQS were performed with a single intermediate image.

Using Eq. (8) in an analogous manner for the 20Ni-10Cr-1P-1Vac quasi-random structure, two binding energies can be calculated by

$$\begin{aligned} E^{\mathrm{P}}_{\mathrm{vb},1} = {} & E\left(20\mathrm{Ni}+10\mathrm{Cr}+1\mathrm{P}+1\mathrm{v}\right)_{\Omega} - E\left(20\mathrm{Ni}+11\mathrm{Cr}+1\mathrm{P}\right)_{\Omega'} \\ & -E\left(21\mathrm{Ni}+10\mathrm{Cr}+1\mathrm{v}\right)_{\Omega} + E\left(21\mathrm{Ni}+11\mathrm{Cr}\right)_{\Omega'} \end{aligned} \tag{12}$$

$$\begin{aligned} E^{\mathrm{P}}_{\mathrm{vb},2} = {} & E\left(20\mathrm{Ni}+10\mathrm{Cr}+1\mathrm{P}+1\mathrm{v}\right)_{\Omega} - E\left(21\mathrm{Ni}+10\mathrm{Cr}+1\mathrm{P}\right)_{\Omega'} \\ & -E\left(20\mathrm{Ni}+11\mathrm{Cr}+1\mathrm{v}\right)_{\Omega} + E\left(21\mathrm{Ni}+11\mathrm{Cr}\right)_{\Omega'} \end{aligned} \tag{13}$$

which give −0.24 eV and −0.39 eV, respectively. These values are similar to that of fcc Ni with dilute phosphorus (−0.29 eV) showing pronounced vacancy binding. Figure 10(b) shows the migration barriers of the four rotation ($\omega_1$-type) hops and one exchange ($\omega_2$-type) hops with and without the P atom. The mean migration barriers of all $\omega_1$- and $\omega_2$-type hops with a P atom in the SQS are ~0.1-0.3 eV lower than those without P (either replaced by Ni or Cr), depending on local configuration. These results suggest a similar phosphorus effect on vacancy kinetics in fcc Ni (Section 4.2), that phosphorus-vacancy pair diffusion can be a more efficient diffusion mechanism compared to the transport by free vacancies, and vacancy-mediated processes like solvent diffusion and LRO can be accelerated accordingly by a dilute addition of phosphorus.

Note that the calculations for SQS in this section consider a single quasi-random structure, so the effects of diffusion correlation and the evolution of local atomic configurations such as that during short-range or long-range ordering were not taken into account. Rigorous and robust simulations of vacancy-mediated processes in concentrated alloys would require more advanced methods like kinetic Monte Carlo modeling with accurate energetics constructed by cluster expansion and DFT methods [69,70], but the techniques would increase the model complexity significantly for multi-component (Ni-Cr-P-Vacancy) consideration and are thus outside the scope of this work. Nonetheless, the DFT calculations for 20Ni-10Cr-1P-1Vac SQS and fcc Ni clearly show the similarities of the pronounced phosphorus effects: generating strong vacancy binding with low exchange barrier and reducing the migration barrier of rotation hops, which together enhance vacancy kinetics activated by the mobile vacancy-phosphorus pairs. These similarities rationalize the use of fcc Ni as a reasonable approximation to understand the phosphorus-enhanced vacancy-mediated processes in concentrated Ni-Cr alloys. The fundamental concept of phosphorus effects may also apply to industrial Ni-Cr alloys like Alloy 690 with ~30 wt.% Cr, but accurate and quantitative predictions of such effects would be challenging because of the complex solute synergies, such as the effect of non-dilute Fe (~10 wt.% in Alloy 690) on thermo-kinetic properties. The addition of Fe was found to reduce the order-disorder transition temperature and thermodynamic driving force of $Ni_2Cr$ ordering [6,7,88], and hence it might in some degree counteract the acceleration effect of phosphorus. Cluster expansion and DFT methods would be valuable tools to characterize these synergetic interactions and pave the way to construct accurate energetics for complex multi-component alloy system.

## 6. Conclusions

We have performed *ab initio* first-principle DFT calculations of phosphorus-enhanced diffusion of solvent atoms mediated by vacancies in fcc Ni. The diffusion calculation is based on the five-frequency model taking into account the effects of vacancy-solute interaction, free and bounded vacancies, as well as correlation of solute-vacancy hops. The DFT calculation shows pronounced vacancy-solute binding (−0.29 eV), which is predominantly caused by the chemical effect rather than the size misfit. Additionally, CI-NEB calculations show that the migration barriers of both vacancy-solute exchange ($\omega_2$) and solvent rotation ($\omega_1$) hops are at least 0.35 eV lower than that of vacancy-atom exchange in pure solvent ($\omega_0$). These results indicate an efficient solute diffusion mechanism operated by solute-vacancy pairs, which also enhance the vacancy mobility and solvent diffusion wherever the phosphorus atoms hop around. The calculated activation energy of P tracer diffusion in dilute Ni-P is 2.16 eV, which is ~0.7 eV lower than Ni self-diffusion. We further investigated the solute-enhanced solvent and vacancy diffusion and demonstrated that the increase in phosphorus composition and decrease in temperature shift the dominant vacancy kinetics from free (unbounded) vacancies to vacancy-solute pairs (bounded). The model predicts more than 10 times acceleration of vacancy-mediated solvent diffusion when the phosphorus composition is not well restricted below 0.1 at.% at a temperature lower than 600°C. The predictions are consistent with the unexpected phosphorus-accelerated ordering transformation as reported by early studies.

The model calculations are accompanied by experiments showing the acceleration of $Ni_2Cr$ ordering transformation by alloying minor amounts of phosphorus from 0.01 to 0.1 at.% at 373, 418, and 470°C. The ordering kinetics were characterized by tracking the evolution of micro-hardness increase associated with the formation of ordered precipitates, and Dienes kinetic theory of ordering was utilized to calibrate the hardness data at various temperatures and for alloying phosphorus with 0.01, 0.05, 0.075, and 0.1 at.%. We showed that the time evolution of hardness increase can be described well by adjusting the activation energy from 2.26 eV (for alloys without P) to 1.57 eV for the alloy with 0.1 at.% P.

Comparison to other typical 3*sp* elements displays strong resemblance between phosphorus and sulfur, both showing pronounced vacancy binding and reduced $\omega_1$ and $\omega_2$ migration barriers. These characteristic can be correlated with the pronounced sharing of electrons between the solute atom and Ni either at as initial ground state or transition state, as manifested by the electron charge density contour and Bader charge analysis. It is thus expected that sulfur can cause significant enhancement of solvent diffusion as well if the impurity is not well controlled below 0.01 at.%.

The overall results demonstrate a critical role of phosphorus in modifying vacancy-mediated kinetics in fcc Ni alloys, including solvent diffusion and ordering transformations. DFT and CI-NEB calculations for fcc Ni and the quasi-random structure of Ni-33 at.% Cr with P manifest strong similarities in terms of pronounced phosphorus-vacancy binding and low migration barriers of exchange and rotation hops, together enhancing vacancy mobility. It has commonly been known that solutes with strong vacancy binding can diminish vacancy-mediated ordering kinetics through the operation of solute trapping. The present work clarifies that strong vacancy-binding solutes like phosphorus can significantly increase vacancy mobility mediated by bounded or paired vacancies with solute atoms. This phosphorus-enhanced vacancy kinetics was also observed experimentally in Fe-based austenitic steels. Such an effect may influence rate processes dramatically, cause variability of microstructure changes in industrial alloys and increase the risk to cast or welded components where solidification segregation of phosphorus or other impurities locally increases concentrations.

**Acknowledgements**

This material is based upon work supported by the National Science Foundation under Grant No. 1653123-DMR. This research is being performed using funding received from the DOE Office of Nuclear Energy's Nuclear Energy University Program, Cooperative Agreement Number DE-NE0008423.

## FIGURE CAPTION

Figure 1. Schematic plot showing the five-frequency model for dilute fcc solid solutions. (Figure modified from ref. [38])

Figure 2. Plots of migration barriers showing the minimum energy paths of the five distinctive types of vacancy-atom exchanges in a Ni-P dilute alloy obtained by the CI-NEB method. The atomic hops are described by the five-frequency model as displayed in Figure 1.

Figure 3. Plots showing the temperature dependence of Ni self-diffusion coefficient in pure fcc Ni and P tracer diffusion coefficient in dilute Ni-P.

Figure 4. Plots showing the (a) temperature dependence of Ni solvent diffusion coefficient in Ni-P alloys with various solute compositions from 0 to 1 at.%. (b) enhancement of Ni solvent diffusion as a function of P solute concentration at various temperature from 300 to 700°C.

Figure 5. Plots showing the temperature dependence of (a) the effective vacancy diffusion coefficient and (b) the fraction of vacancy diffusion by solute-vacancy pairs for dilute Ni-P alloys with various P compositions (at.%). Also shown in (a) are the two vacancy diffusion limits by free vacancies and vacancy-solute pairs, which are plotted by ○ and △ symbols, respectively.

Figure 6. Plot showing the acceleration factor on time to reach a 50 HV increase of micro-hardness during aging at 470 °C as a function of phosphorus concentration (at.%) in Ni-33at.%Cr alloys. The solid line is the linear fit of data.

Figure 7. Plots showing the fraction of hardness increase as a function of isothermal aging time for Ni-33%Cr alloys with 0, 0.01, 0.05, 0.075, and 0.1 at.% phosphorus at (a) 373 °C, (b) 418 °C, and (c) 470 °C. The symbols show the data from experiment and the line plots show the result of Dienes ordering model.

Figure 8. Contour maps showing the electron charge density (in a unit of $Bohr^{-3}$) in the (111) plane of fcc Ni containing (a) 107 Ni atoms and one phosphorus atom and (b) 106 Ni atoms and one phosphorus atom with a vacancy in nearest neighbor (a vacancy-phosphorus pair). The (111) plane is selected to include (a) the phosphorus atom and (b) vacancy-phosphorus pair. The number below the element name represent the Bader charge of each atom in a unit of *e*. Bader charges in magnitude smaller than 0.01 are not shown.

Figure 9. Contour maps showing the electron charge density (in a unit of $Bohr^{-3}$) in the (111) plane of fcc Ni at the transition state of the (a) $\omega_1$ rotation hop and (b) $\omega_2$ exchange hop as defined by the five-frequency model (Section 2.1). The migrating atoms (marked by bold font) are at the saddle point of their migration path and can be identified as the atom in between two charge density concave wells. The number below the element name represent the Bader charge of each atom in a unit of *e*. Bader charges in magnitude smaller than 0.01 are not shown.

Figure 10. (a) Schematic plot showing the special quasi-random structure (SQS) of the 2×2×2 fcc supercell consisting of 20 Ni atoms, 10 Cr atoms, 1 P atom, and 1 vacancy. The Ni, Cr, and P atoms are displayed as blue, green, and purple spheres, and the vacancy is the white square labeled as “Vac”. The four spheres labeled by the element name with number are the atoms in the nearest-neighbor distance with the P atom and vacancy. (b)

Plot showing the calculated migration barriers of the four rotation ($\omega_1$-type) hops and one exchange ($\omega_2$-type) hop with phosphorus (purple) and without phosphorus (blue and green, each being replaced by Ni and Cr, respectively). Each hop corresponds to the exchange of vacancy and the labeled atom shown in (a). The "error bar" shows the range of migration barrier due to forward and backward hops, while the square shows the mean migration barrier of each migration path.

Table 1. Thermo-kinetic parameters of pure Ni and Ni-P dilute alloy calculated by the first-principles DFT method in the present study.

| Parameter | Value | unit |
|---|---|---|
| *Ni parameters:* | | |
| Lattice constant of pure Ni | 0.352 | nm |
| Migration barrier $E_0$ | 1.06 | eV |
| Attempt frequency $\nu_{hop}^{Ni}$ | $4.63\times10^{12}$ | Hz |
| *P in Ni parameters:* | | |
| P-Vacancy binding energy $E_{vb}$ | −0.30 (Eq.(3))<br>−0.29 (Eq.(8)) | eV<br>eV |
| Migration barrier $E_1$ | 0.70 | eV |
| Migration barrier $E_2$ | 0.62 | eV |
| Migration barrier $E_{3\text{-}2nn}$ | 1.30 | eV |
| Migration barrier $E_{3\text{-}3nn}$ | 1.13 | eV |
| Migration barrier $E_{3\text{-}4nn}$ | 1.13 | eV |
| Migration barrier $E_{4\text{-}2nn}$ | 1.04 | eV |
| Migration barrier $E_{4\text{-}3nn}$ | 0.83 | eV |
| Migration barrier $E_{4\text{-}4nn}$ | 0.84 | eV |
| Attempt frequency $\nu_{hop}^{P}$ | $3.32\times10^{12}$ | Hz |

Table 2. Micro-hardness data showing the change of hardness for the five experimental Ni-33at%Cr alloys with phosphorus concentration from 0 to 0.1 at.% after aging at 373, 418, and 470 °C. The micro-hardness of the as-received alloys are set as reference and the values in the bracket are measured hardness.

| Temperature (°C) | Aging time (h) | ΔH 0 at.%P | ΔH 0.01at%P | ΔH 0.05at%P | ΔH 0.075at%P | ΔH 0.1at%P |
|---|---|---|---|---|---|---|
| As-received | 0 | 0 (143.0 ± 3.5) | 0 (147.8 ± 6.1) | 0 (143.4 ± 4.9) | 0 (140.0 ± 4.8) | 0 (144.6 ± 2.9) |
| 373 | 10 | - | - | - | - | 115.6 |
| 373 | 30 | 5.2 | 10 | 69.8 | 83.2 | - |
| 373 | 126.5 | 3.2 | 4.4 | 81.4 | 117.4 | 127 |
| 373 | 300 | -2.8 | 33 | 79.8 | 130.6 | 136 |
| 373 | 1000 | 10.4 | 58.4 | 120 | 132.4 | 133.2 |
| 373 | 3000 | 43.2 | 82 | 124 | 137.4 | 128.4 |
| 418 | 3 | - | - | - | 70.2 | 23.4 |
| 418 | 10 | 5.2 | 10.6 | 65 | - | - |
| 418 | 30 | −7.4 | 14.2 | 82.8 | 116.8 | 126.2 |
| 418 | 100 | 5.6 | 65.4 | 108 | 144 | 135.8 |
| 418 | 300 | 64.2 | 88 | 111.8 | 127 | 138.4 |
| 418 | 1000 | 89.8 | 108 | 155 | 158.4 | 155.4 |
| 418 | 3000 | 89.6 | 123.2 | 132 | 157.2 | 143.8 |
| 470 | 1 | 8.2 | 4 | 60.2 | 77.6 | 110.8 |
| 470 | 10 | 12.2 | 36.2 | 99.2 | 118.4 | 136.8 |
| 470 | 30 | 46.8 | 66.4 | 121.4 | 132.6 | 140.6 |
| 470 | 100 | 86.2 | 110 | 122.8 | 138.8 | 144.6 |
| 470 | 300 | 104 | 110.2 | 140.8 | 146.2 | 142 |
| 470 | 1000 | 107 | 123 | 148.8 | 151.8 | 151 |
| 470 | 3000 | 112.4 | 121.8 | 131.4 | 152.6 | 155 |

Table 3. Parameters used in the kinetic ordering model. The variables marked with a * are activation energies that were adjusted to fit with the hardness data for each alloy in the present study.

| Parameter | Value | unit |
|---|---|---|
| Activation energy *U* (0% P) | 2.26* | eV |
| | 2.14 [61] | eV |
| Activation energy *U* (0.01% P) | 2.18* | eV |
| Activation energy *U* (0.05% P) | 2.01* | eV |
| Activation energy *U* (0.075% P) | 1.83* | eV |
| Activation energy *U* (0.1% P) | 1.57* | eV |
| Ordering energy $W_0$ | 0.35 [61] | eV |
| Concentration at stoichiometry $\xi$ | 0.33 [61] | |
| Vibration frequency | $10^{13}$ [61] | Hz |
| Ordering fluctuation | $10^{-3}$ | |

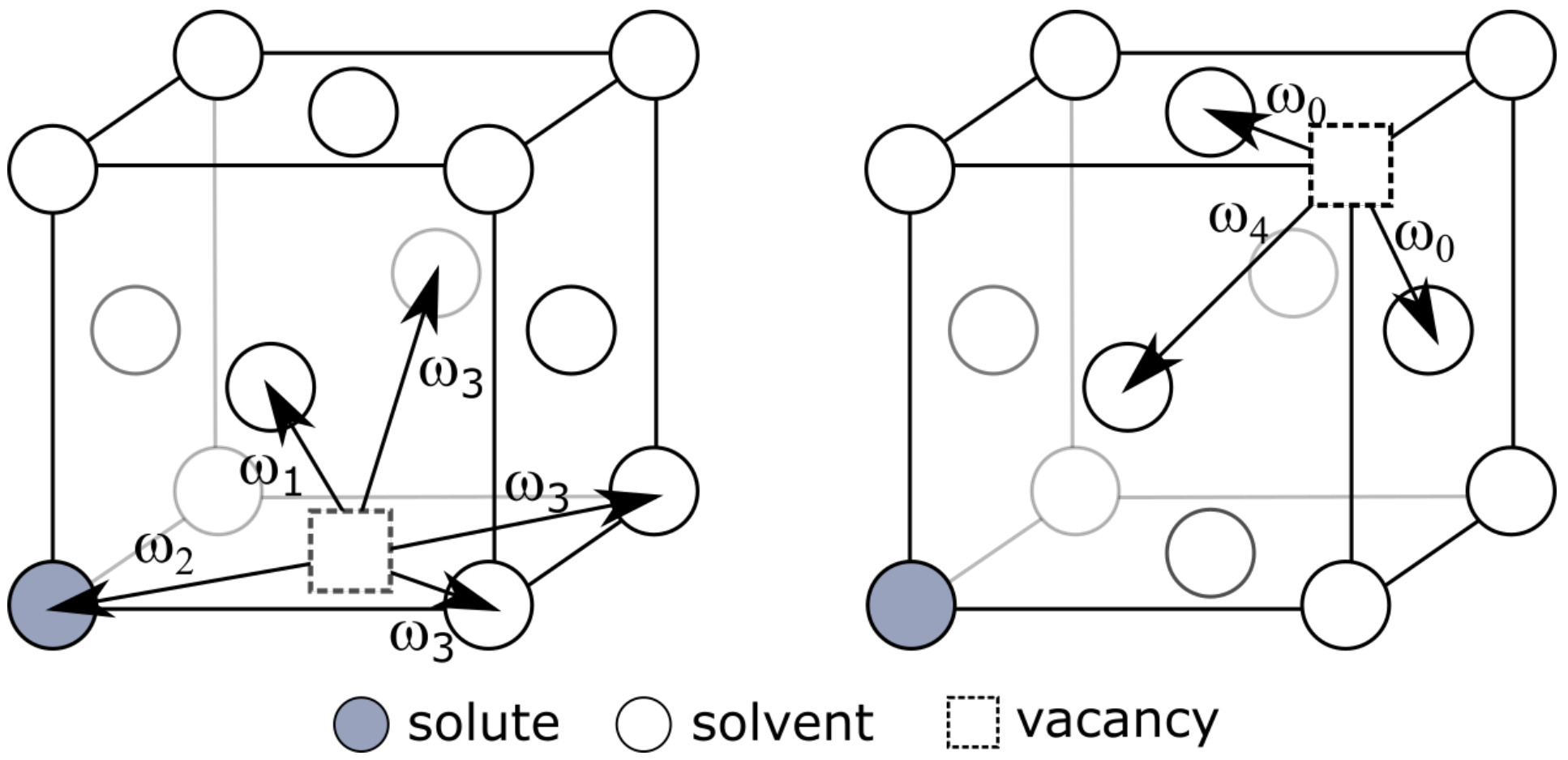

Figure 1. Schematic plot showing the five-frequency model for dilute fcc solid solutions. (Figure modified from ref. [38])

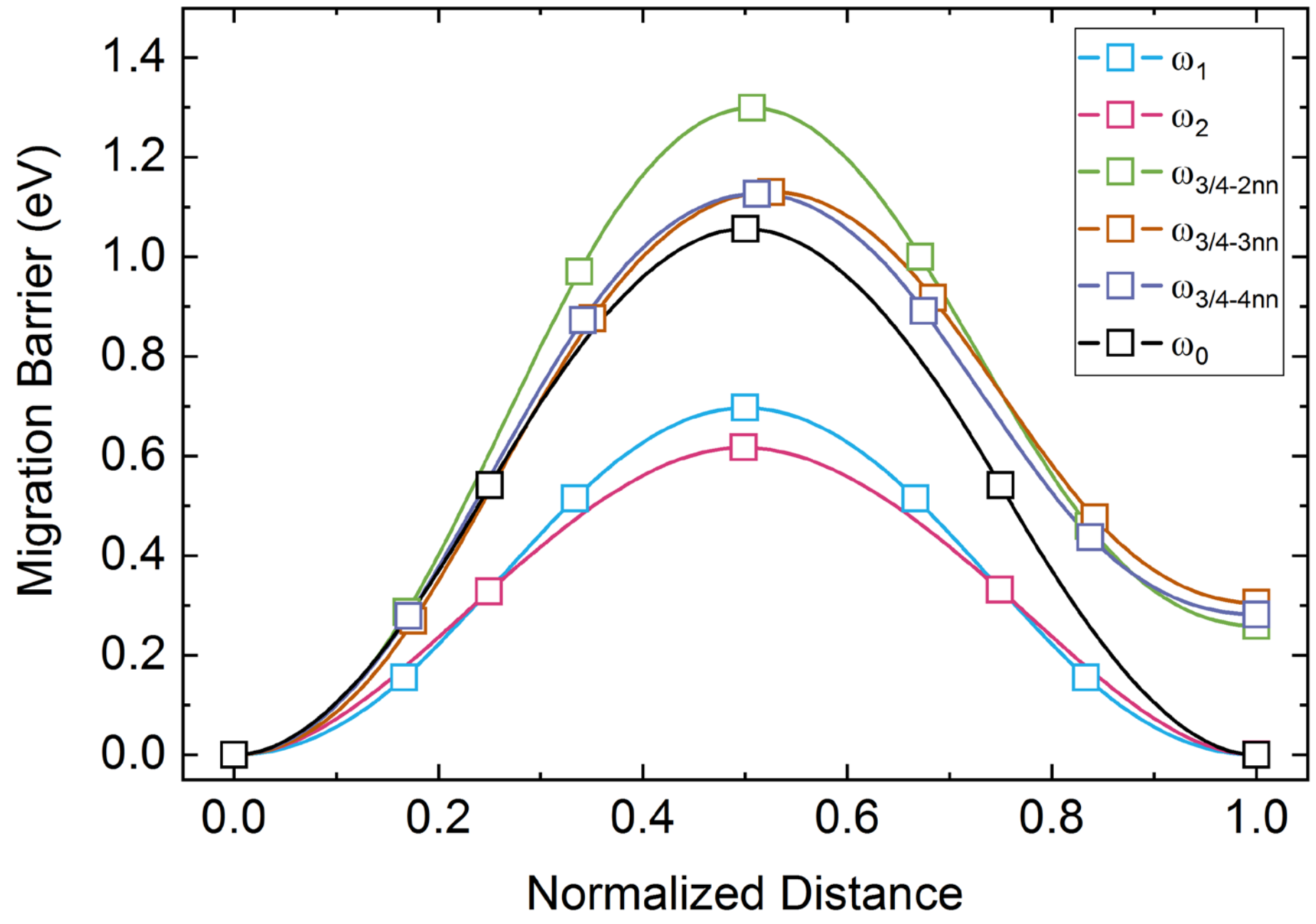

Figure 2. Plots of migration barriers showing the minimum energy paths of the five distinctive types of vacancy-atom exchanges in a Ni-P dilute alloy obtained by the CI-NEB method. The atomic hops are described by the five-frequency model as displayed in Figure 1.

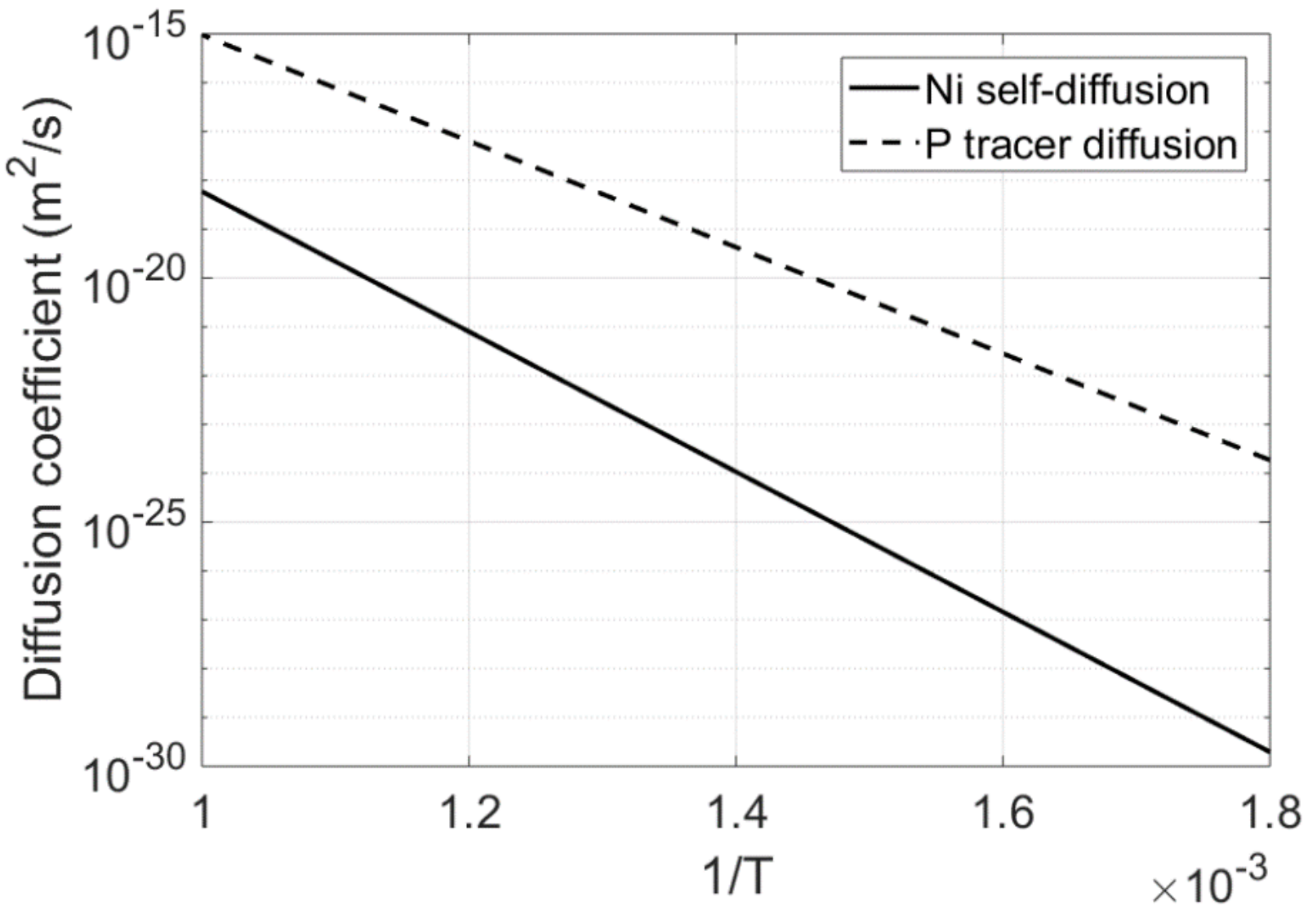

Figure 3. Plots showing the temperature dependence of Ni self-diffusion coefficient in pure fcc Ni and P tracer diffusion coefficient in dilute Ni-P.

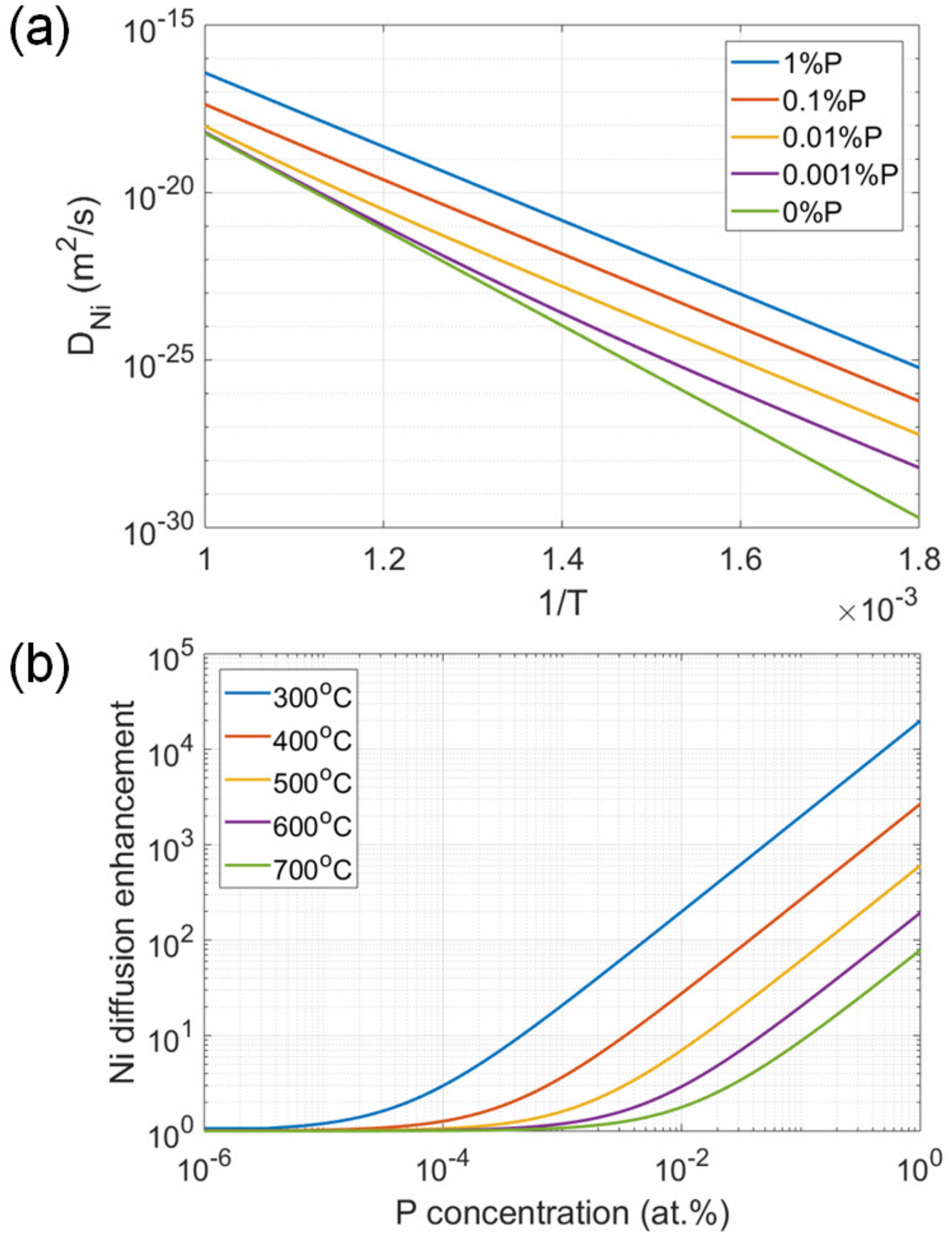

Figure 4. Plots showing the (a) temperature dependence of Ni solvent diffusion coefficient in Ni-P alloys with various solute compositions from 0 to 1 at.%. (b) enhancement of Ni solvent diffusion as a function of P solute concentration at various temperature from 300 to 700°C.

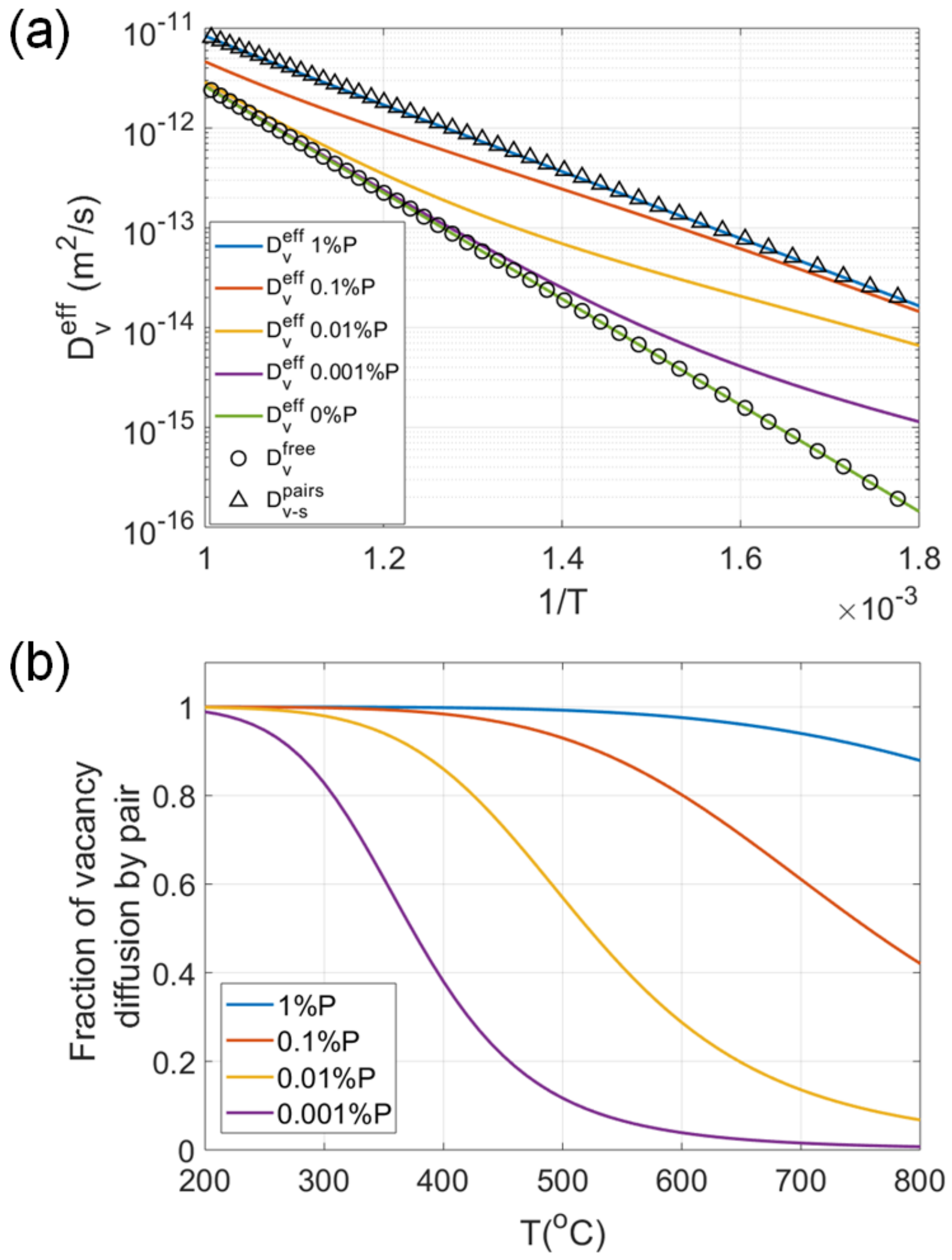

Figure 5. Plots showing the temperature dependence of (a) the effective vacancy diffusion coefficient and (b) the fraction of vacancy diffusion by solute-vacancy pairs for dilute Ni-P alloys with various P compositions (at.%). Also shown in (a) are the two vacancy diffusion limits by free vacancies and vacancy-solute pairs, which are plotted by ○ and △ symbols, respectively.

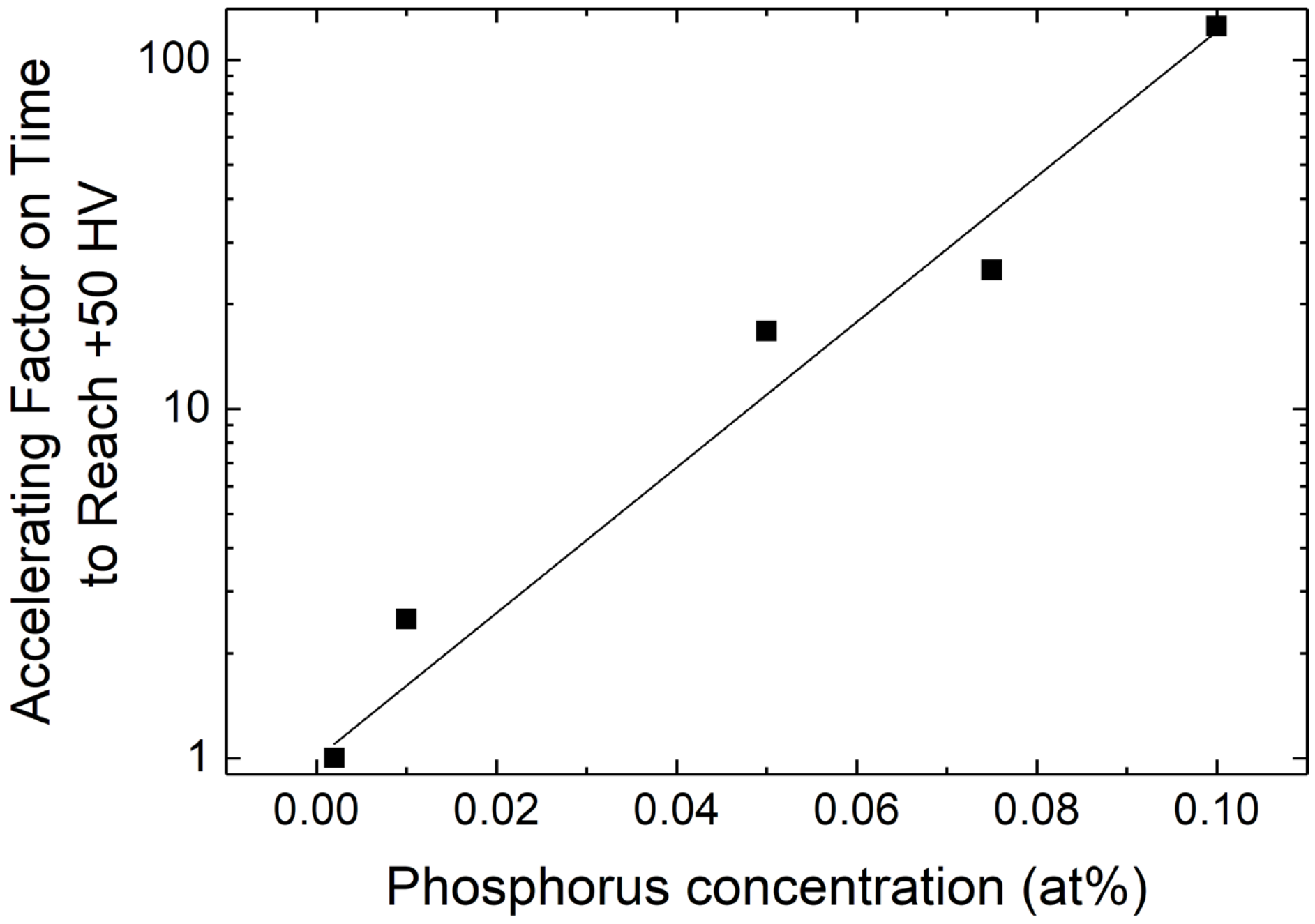

Figure 6. Plot showing the acceleration factor on time to reach a 50 HV increase of micro-hardness during aging at 470 °C as a function of phosphorus concentration (at.%) in Ni-33at.%Cr alloys. The solid line is the linear fit of data.

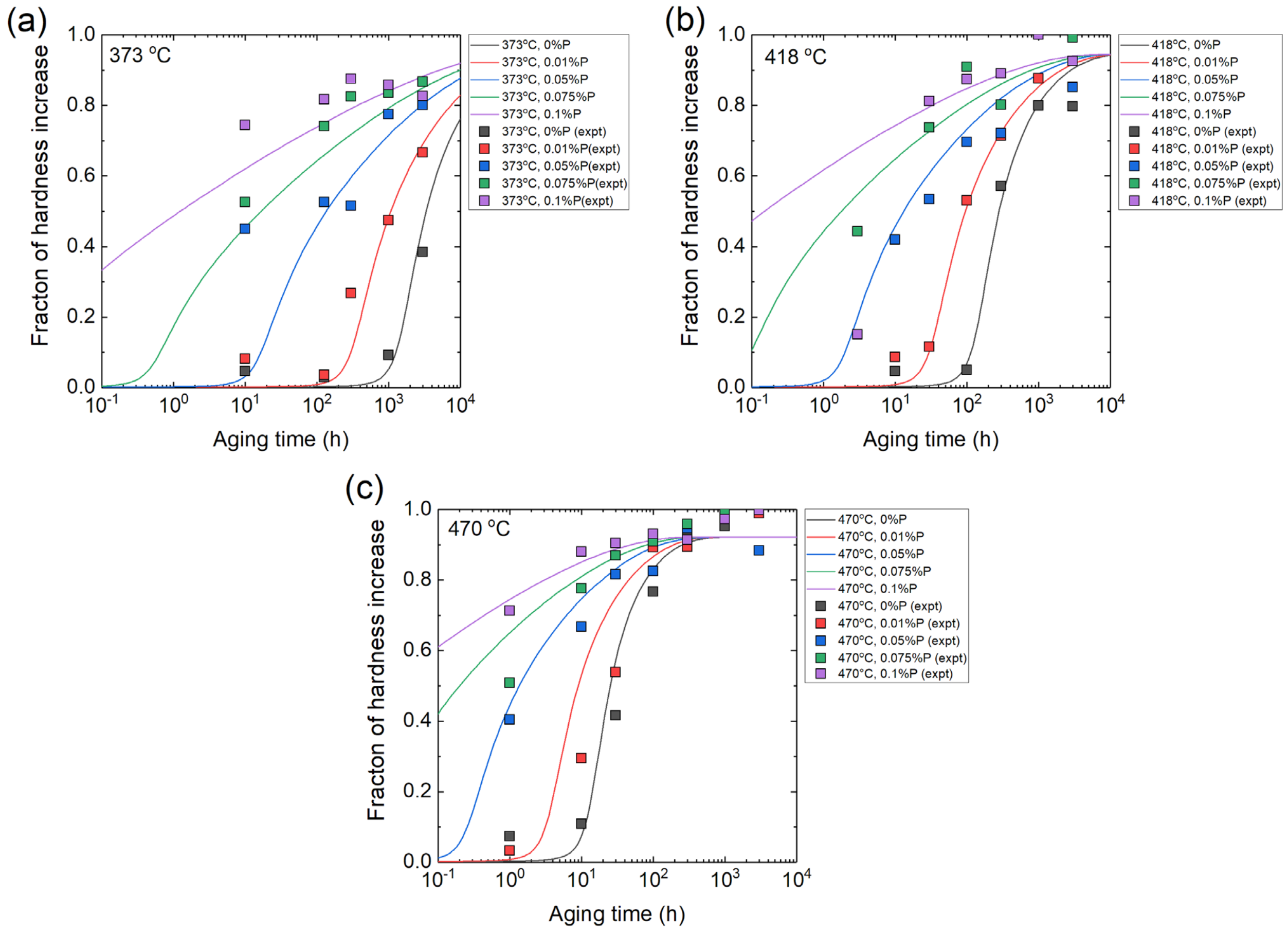

Figure 7. Plots showing the fraction of hardness increase as a function of isothermal aging time for Ni-33%Cr alloys with 0, 0.01, 0.05, 0.075, and 0.1 at.% phosphorus at (a) 373 °C, (b) 418 °C, and (c) 470 °C. The symbols show the data from experiment and the line plots show the result of Dienes ordering model.

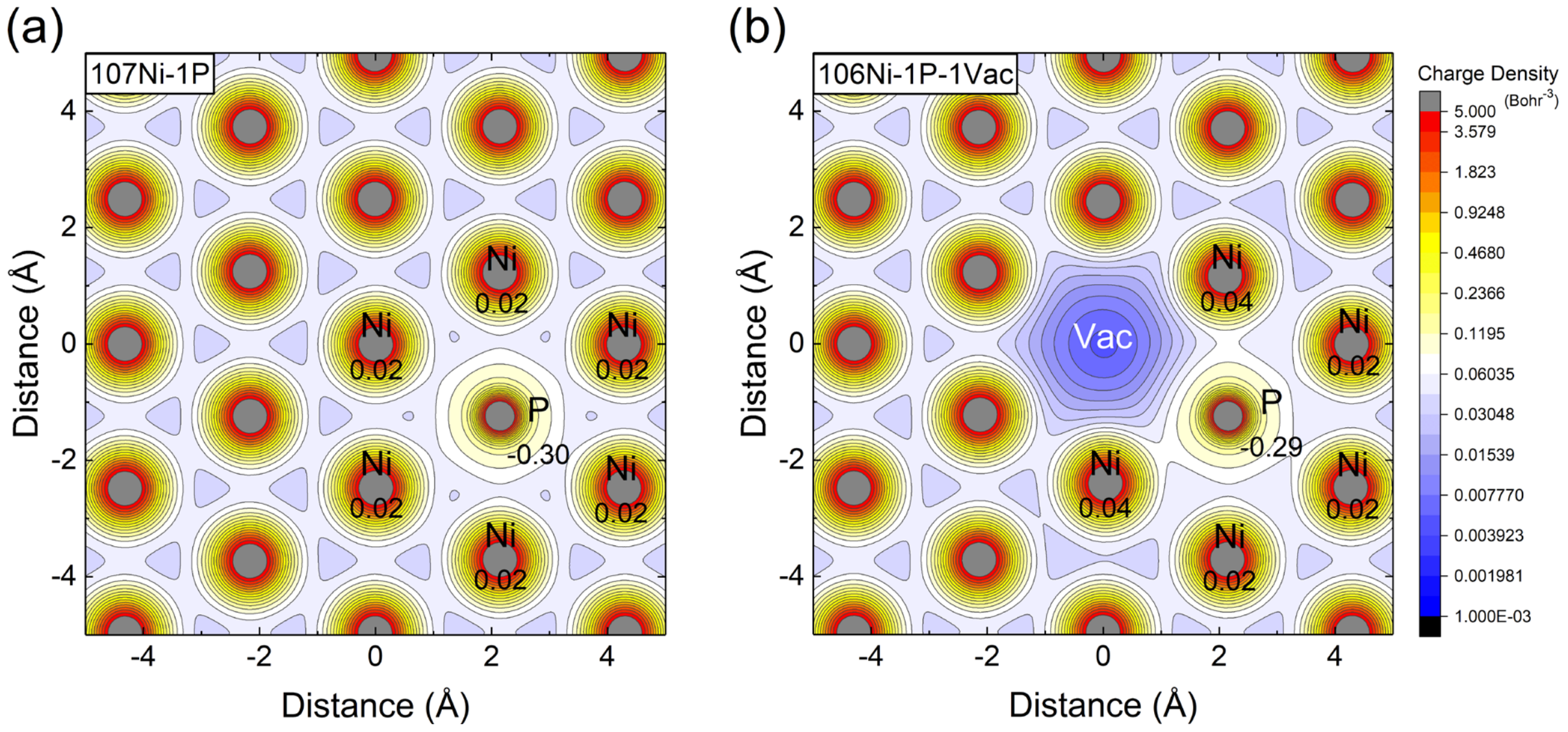

Figure 8. Contour maps showing the electron charge density (in a unit of $Bohr^{-3}$) in the (111) plane of fcc Ni containing (a) 107 Ni atoms and one phosphorus atom and (b) 106 Ni atoms and one phosphorus atom with a vacancy in nearest neighbor (a vacancy-phosphorus pair). The (111) plane is selected to include (a) the phosphorus atom and (b) vacancy-phosphorus pair. The number below the element name represent the Bader charge of each atom in a unit of *e*. Bader charges in magnitude smaller than 0.01 are not shown.

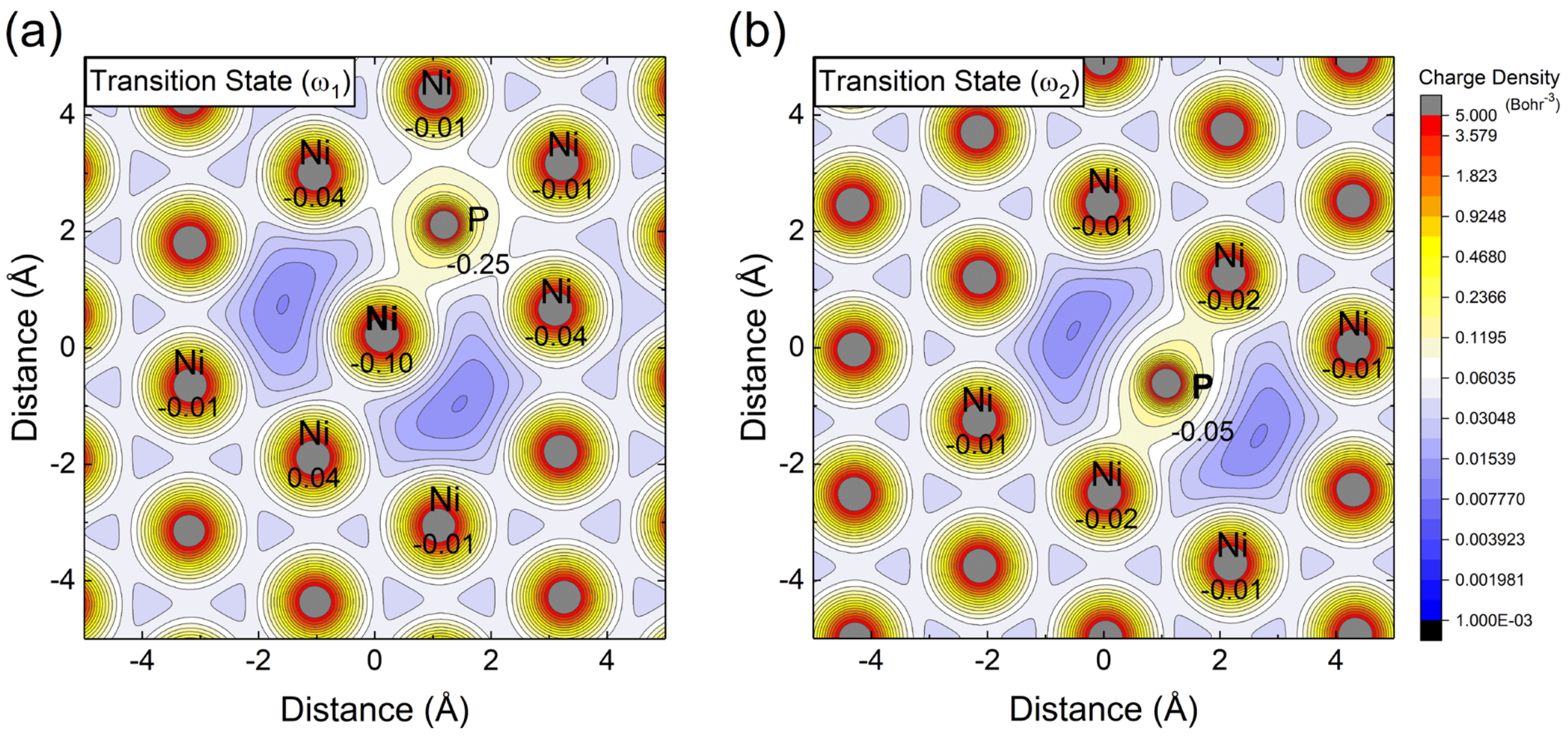

Figure 9. Contour maps showing the electron charge density (in a unit of Bohr$^{-3}$) in the (111) plane of fcc Ni at the transition state of the (a) $\omega_1$ rotation hop and (b) $\omega_2$ exchange hop as defined by the five-frequency model (Section 2.1). The migrating atoms (marked by bold font) are at the saddle point of their migration path and can be identified as the atom in between two charge density concave wells. The number below the element name represent the Bader charge of each atom in a unit of *e*. Bader charges in magnitude smaller than 0.01 are not shown.

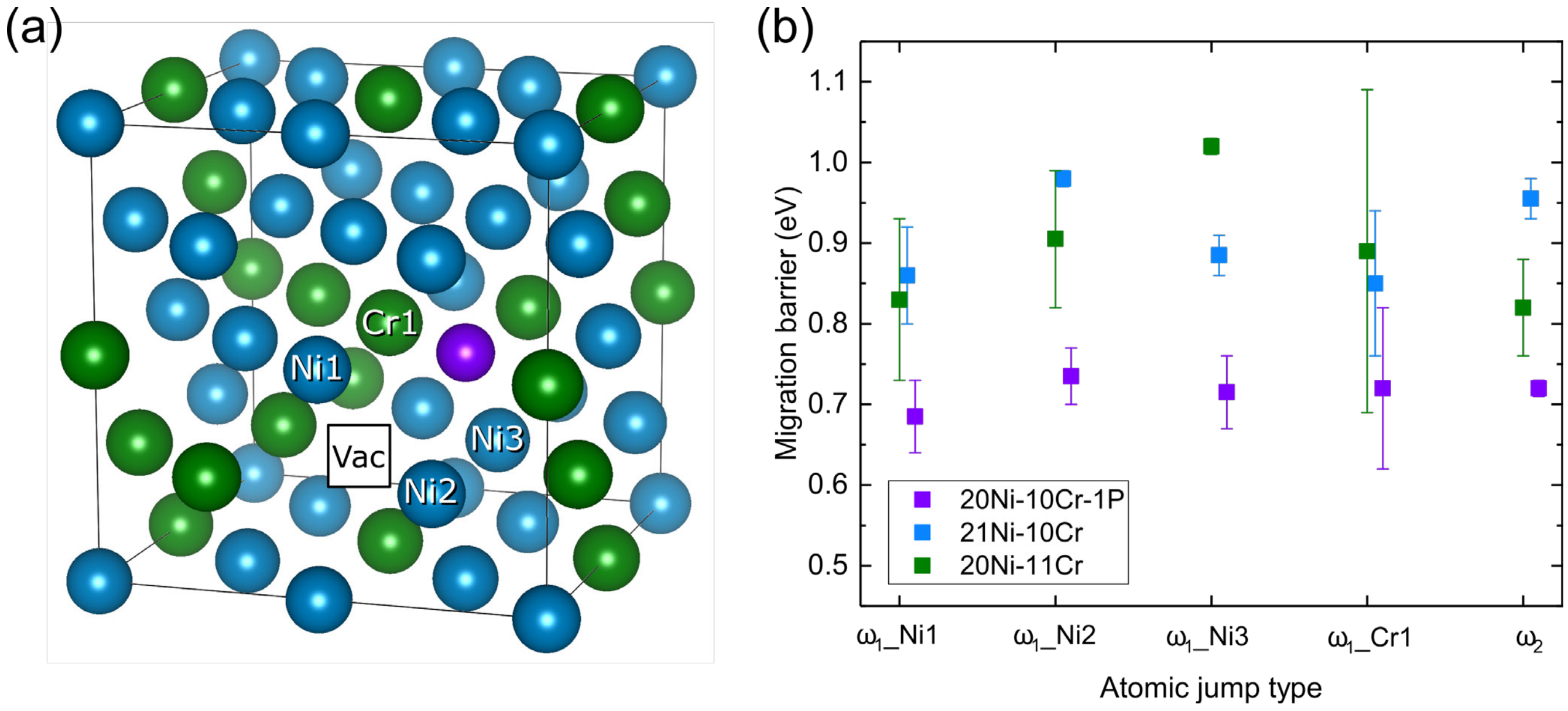

Figure 10. (a) Schematic plot showing the special quasi-random structure (SQS) of the 2×2×2 fcc supercell consisting of 20 Ni atoms, 10 Cr atoms, 1 P atom, and 1 vacancy. The Ni, Cr, and P atoms are displayed as blue, green, and purple spheres, and the vacancy is the white square labeled as "Vac". The four spheres labeled by the element name with number are the atoms in the nearest-neighbor distance with the P atom and vacancy. (b) Plot showing the calculated migration barriers of the four rotation ($\omega_1$-type) hops and one exchange ($\omega_2$-type) hop with phosphorus (purple) and without phosphorus (blue and green, each being replaced by Ni and Cr, respectively). Each hop corresponds to the exchange of vacancy and the labeled atom shown in (a). The "error bar" shows the range of migration barrier due to forward and backward hops, while the square shows the mean migration barrier of each migration path.

**Supplementary Information**

# Ab initio study of phosphorus effect on vacancy-mediated process in nickel alloys – an insight into $Ni_2Cr$ ordering

Jia-Hong Ke[1], George A. Young[1,2], and Julie D. Tucker[1]

[1]Oregon State University, 204 Rogers Hall, Corvallis, OR 97331
[2] Kairos Power, 707 W. Tower Ave., Alameda, CA 94501

## S1. Diffusion coefficients calculations based on multi-frequency model of fcc

We implemented the multi-frequency framework developed by Lidiard and Le Claire [1–3] to calculate the solute effected diffusion properties mediated by vacancies. For dilute fcc-based solid solutions, the diffusion properties can be predicted by the rates of five distinctive types of vacancy-atom exchanges relative to the solute atom. This particular set of rates is known as the five-frequency model as developed by Lidiard and Le Claire [1,2,4]. The model assumes a dilute solution condition and therefore does not consider solute-solute interactions. Figure 1 in the manuscript illustrates the various diffusion events $\omega_i$ of the five-frequency model following ref. [4]. $\omega_0$ is the hop rate of vacancy-atom exchange in the pure solvent away from any solutes, $\omega_1$ is the vacancy exchange with a 1st nearest neighbor of the solute or vacancy-solute rotation hop, $\omega_2$ is vacancy-solute exchange hop, and $\omega_3/\omega_4$ are the vacancy-solute dissociation/association hops from/to a 2nd, 3rd, or 4th nearest-neighbor solvent atom with respect to the solute atom. Each jump frequency in the five-frequency model can be calculated based on the transition state theory [5,6] by the following Arrhenius expression:

$$\omega_i = \nu_i \exp(-E_i / k_B T) \quad \text{(S1)}$$

where $i$ indicates the type of vacancy-atom hops, $\nu_i$ is the attempt frequency for the hop and $E_i$ is the migration barrier which can be calculated from DFT and the CI-NEB method.

In the present study, we assume that the effective attempt frequencies can be determined by using two prefactors, including the one for migration of the solute atom of an isolated vacancy-solute pair and the other for migration of a solvent atom in pure solvent. The effective attempt frequency was calculated based on Vineyard's harmonic transition state theory. We further follow the approximation of Wu *et al.* [7] by considering only the vibration mode of the hopping atom between the initial ground state position ( $\nu_j^{\text{initial}}$ ) and the saddle point transition-state configuration ( $\nu_j^{\text{saddle}}$ ). The attempt frequency can then be estimated as

$$\nu_{\text{hop}} = \frac{\prod_{j=1}^{3n} \nu_j^{\text{initial}}}{\prod_{j=1}^{3n-1} \nu_j^{\text{saddle}}} \sim \frac{\prod_{j=1}^{3} \nu_j^{\text{initial}}}{\prod_{j=1}^{2} \nu_j^{\text{saddle}}} \tag{S2}$$

Previous studies indicated that this approximation is reliable and further enhanced treatments of the phonon mode calculation would not improve the predictions of solute diffusion significantly [7]. Since the diffusion properties are dominated by the energetics of vacancy migration and vacancy-solute interactions, the approximations will not alter the conclusion and contribution of the present work.

The phenomenological coefficients $L_{\text{ij}}$ of atom transport for the five-frequency model were obtained based on the second-shell approximation [3].

$$L_{\text{AA}} = \frac{ns^2}{6k_BT}\left[12c'_{\text{v}}\left(1-7c'_{\text{B}}\right)\omega_0 + c_{\text{pair}}\left(4\omega_1+14\omega_3\right) + c_{\text{pair}}A_{\text{AA}}^{(1)}\right] \tag{S3}$$

$$L_{\text{AB}} = L_{\text{BA}} = \frac{ns^2}{6k_BT}c_{\text{pair}}A_{\text{AB}}^{1} \tag{S4}$$

$$L_{\text{BB}} = \frac{ns^2}{6k_BT}\left(c_{\text{pair}}\omega_2 - 2c_{\text{pair}}\omega_2^{\,2}/\Omega\right) \tag{S5}$$

where

$$\begin{aligned} A_{\text{AA}}^{(1)} = \Big[&-2\left(3\omega_3-2\omega_1\right)^2 + 28\omega_3\left(1-F\right)\left(3\omega_3-2\omega_1\right)\left(\omega_0/\omega_4-1\right) \\ &-14\omega_3\left(1-F\right)\left(2\omega_1+2\omega_2+7\omega_3\right)\left(\omega_0/\omega_4-1\right)^2\Big]\Big/\Omega \end{aligned} \tag{S6}$$

$$A_{\text{AB}}^{(1)} = \omega_2\left[2\left(3\omega_3-2\omega_1\right)+14\omega_3\left(1-F\right)\left(\omega_0/\omega_4-1\right)\right]\big/\Omega \tag{S7}$$

and

$$\Omega = 2\omega_1 + 2\omega_2 + 7\omega_3 F \tag{S8}$$

A and B correspond to the solvent and solute atoms, respectively, $n$ denotes the atomic site number density, and $s$ is the nearest-neighbor spacing ( $\sqrt{2}a_0$ , $a_0$ is the fcc lattice constant). $F$ is a vacancy escape factor as a function of $\omega_4/\omega_0$ which can be deduced under the two-shell approximation. The factor $F$ has been determined by Koiwa and Ishioka as [8]

$$7(1-F)=\frac{10\xi^4+180.3\xi^3+924.3\xi^2+1338.1\xi}{2\xi^4+40.1\xi^3+253.3\xi^2+596.0\xi+435.3} \quad \text{(S9)}$$

where $\xi = \omega_4/\omega_0$. Note that there are three separate vacancy-solute dissociation/association hop frequencies, and a single effective frequency is used for the five-frequency model. The effective vacancy-solute dissociation/association hop frequencies can be determined by assigning the same rate to the seven dissociation/association pathways, which are given by the following equation

$$7\omega_{3/4}=2\omega_{3/4}^{2\text{nn}}+4\omega_{3/4}^{3\text{nn}}+\omega_{3/4}^{4\text{nn}} \quad \text{(S10)}$$

In Eq. (S3)-(S5), $c_{\text{pair}}$ is the site fraction of solute atoms that have a vacancy at their nearest neighbor sites. $c'_{\text{v}}$ and $c'_{\text{B}}$ are the fraction of unbounded vacancies and solute atoms, or $c_{\text{v}}-c_{\text{pair}}$ and $c_{\text{B}}-c_{\text{pair}}$, respectively. In the five-frequency model with dilute and isolated solute atoms, these site fractions can be determined analytically by the conditions of detailed balance and local thermodynamic equilibrium between pairs and unbounded vacancies [3].

$$\frac{c_{\text{pair}}}{12c'_{\text{v}}c'_{\text{B}}}=\exp\left(-\frac{E_{\text{vb}}}{k_BT}\right)=\frac{\omega_4}{\omega_3} \quad \text{(S11)}$$

where $E_{\text{vb}}$ is the binding energy of the first nearest-neighbor vacancy-solute pair.

The tracer diffusion coefficient $D_{\text{B}}$ can be determined by the phenomenological coefficients in the limit of a dilute alloy ($c_{\text{B}}\to 0$) as $k_BT\,L_{BB}/nc_B$ or

$$D_{\text{B}}=2s^2\omega_2c_{\text{v}}f_{\text{B}}\exp\left(-E_{\text{vb}}/k_BT\right) \quad \text{(S12)}$$

where $f_{\text{B}}$ is the correlation factor for solute diffusion which is given by LeClaire and Lidiard as $\left(2\omega_1+7\omega_3F\right)/\left(2\omega_1+2\omega_2+7\omega_3F\right)$ [1]. The self-diffusion of the pure solvent atom can be considered as a special case where B is replaced by an isotope of species A, with equal five frequencies $\omega = \omega_0$ and zero binding energy ($E_{\text{vb}}$). The self-diffusion of the pure solvent atom is given as

$$D_{\text{A}}(0)=2s^2\omega_0c_{\text{v}}f_0 \quad \text{(S13)}$$

where $f_0$ is the correlation factor of solvent self-diffusion and determined as 0.7815 for pure fcc metals.

## S2. Analytical expression of solvent-enhancement factor

In this study we utilized the self-consistent mean field theory developed more recently by Nastar [9] showing better agreement with Monte Carlo simulations than most of the previous models. Under the approximation restricted to nearest-neighbor effective interactions, the model offers an analytical form of the enhancement factor $b$ as a function of the five frequencies [9]

$$b = b_{\omega} + b_f \tag{S14}$$

$$b_{\omega} = -18 + 14\frac{\omega_4}{\omega_0} + 4\frac{\omega_1\omega_4}{\omega_0\omega_3} \tag{S15}$$

$$b_f = -2(1-f_0)\frac{\omega_4}{\omega_0}\frac{\frac{\omega_2}{\omega_1}\left(-27+18\frac{\omega_1}{\omega_3}\right)+\frac{1}{9}\left(164\frac{\omega_1}{\omega_3}+389\frac{\omega_3}{\omega_1}-472\right)}{9\frac{\omega_2}{\omega_1}+7\frac{\omega_3}{\omega_1}+2} \tag{S16}$$

Note that $b_{\omega}$ characterizes the frequency enhancement of the solvent atoms and $b_f$ describes the correlation enhancement.

## S3. Summary of micro-hardness data of five Ni-33at%Cr-P alloys

Table S1. Micro-hardness data showing the effect of aging at 373, 418, and 470 °C on the five experimental Ni-33at%Cr alloys with phosphorus concentration from 0 to 0.1 at.%.

| Temperature (°C) | Aging time (h) | Hardness H (HV 500 g-f) | ΔH (HV 500 g-f) | Aging time (h) | Hardness H (HV 500 g-f) | ΔH (HV 500 g-f) |
|---|---|---|---|---|---|---|
| | **Ni-33at.%Cr** | | | **Ni-33at.%Cr-0.01at%P** | | |
| As-received | 0 | 143.0 ± 3.5 | 0 | 0 | 147.8 ± 6.1 | 0.0 |
| 373 | 30 | 148.2 ± 2.8 | 5.2 | 30 | 157.8 ± 5.5 | 10.0 |
| 373 | 126.5 | 146.2 ± 4.0 | 3.2 | 126.5 | 152.2 ± 4.3 | 4.4 |
| 373 | 300 | 140.2 ± 3.6 | -2.8 | 300 | 180.8 ± 3.9 | 33.0 |
| 373 | 1000 | 153.4 ± 6.2 | 10.4 | 1000 | 206.2 ± 7.9 | 58.4 |
| 373 | 3000 | 186.2 ± 4.9 | 43.2 | 3000 | 229.8 ± 8.9 | 82.0 |
| 418 | 10 | 148.2 ± 3.7 | 5.2 | 10 | 158.4 ± 3.8 | 10.6 |
| 418 | 30 | 135.6 ± 5.2 | -7.4 | 30 | 162.0 ± 5.0 | 14.2 |
| 418 | 100 | 148.6 ± 1.1 | 5.6 | 100 | 213.2 ± 3.9 | 65.4 |
| 418 | 300 | 207.2 ± 9.9 | 64.2 | 300 | 235.8 ± 5.6 | 88.0 |
| 418 | 1000 | 232.8 ± 9.7 | 89.8 | 1000 | 255.8 ± 7.1 | 108.0 |
| 418 | 3000 | 232.6 ± 8.0 | 89.6 | 3000 | 271.0 ± 6.7 | 123.2 |
| 470 | 1 | 151.2 ± 4.0 | 8.2 | 1 | 151.8 ± 4.8 | 4.0 |
| 470 | 10 | 155.2 ± 4.9 | 12.2 | 10 | 184.0 ± 4.5 | 36.2 |
| 470 | 30 | 189.8 ± 12.4 | 46.8 | 30 | 214.2 ± 13.8 | 66.4 |
| 470 | 100 | 229.2 ± 10.3 | 86.2 | 100 | 257.8 ± 5.0 | 110.0 |
| 470 | 300 | 247.0 ± 7.4 | 104.0 | 300 | 258.0 ± 9.6 | 110.2 |
| 470 | 1000 | 250.0 ± 7.3 | 107.0 | 1000 | 270.8 ± 2.9 | 123.0 |
| 470 | 3000 | 255.4 ± 4.7 | 112.4 | 3000 | 269.6 ± 5.1 | 121.8 |
| | **Ni-33at.%Cr-0.05at%P** | | | **Ni-33at.%Cr-0.075at%P** | | |

| | | | | | | | |
|---|---|---|---|---|---|---|---|
| As-received | 0 | 143.4 ± 4.9 | 0.0 | | 0 | 140.0 ± 4.8 | 0.0 |
| 373 | 30 | 213.2 ± 5.0 | 69.8 | | 30 | 223.2 ± 10.6 | 83.2 |
| 373 | 126.5 | 224.8 ± 3.1 | 81.4 | | 126.5 | 257.4 ± 16.6 | 117.4 |
| 373 | 300 | 223.2 ± 3.3 | 79.8 | | 300 | 270.6 ± 12.8 | 130.6 |
| 373 | 1000 | 263.4 ± 18.1 | 120.0 | | 1000 | 272.4 ± 5.3 | 132.4 |
| 373 | 3000 | 267.4 ± 9.9 | 124.0 | | 3000 | 277.4 ± 12.6 | 137.4 |
| 418 | 10 | 208.4 ± 9.4 | 65.0 | | 3 | 210.2 ± 7.5 | 70.2 |
| 418 | 30 | 226.2 ± 7.9 | 82.8 | | 30 | 256.8 ± 5.8 | 116.8 |
| 418 | 100 | 251.4 ± 14.1 | 108.0 | | 100 | 284.0 ± 9.5 | 144.0 |
| 418 | 300 | 255.2 ± 5.8 | 111.8 | | 300 | 267.0 ± 12.6 | 127.0 |
| 418 | 1000 | 298.4 ± 11.9 | 155.0 | | 1000 | 298.4 ± 11.3 | 158.4 |
| 418 | 3000 | 275.4 ± 8.4 | 132.0 | | 3000 | 297.2 ± 4.1 | 157.2 |
| 470 | 1 | 203.6 ± 11.6 | 60.2 | | 1 | 217.6 ± | 77.6 |
| 470 | 10 | 242.6 ± 6.1 | 99.2 | | 10 | 258.4 ± 7.0 | 118.4 |
| 470 | 30 | 264.8 ± 9.9 | 121.4 | | 30 | 272.6 ± 9.7 | 132.6 |
| 470 | 100 | 266.2 ± 10.8 | 122.8 | | 100 | 278.8 ± 2.4 | 138.8 |
| 470 | 300 | 284.2 ± 13.5 | 140.8 | | 300 | 286.2 ± 5.7 | 146.2 |
| 470 | 1000 | 292.2 ± 7.3 | 148.8 | | 1000 | 291.8 ± 8.9 | 151.8 |
| 470 | 3000 | 274.8 ± 6.3 | 131.4 | | 3000 | 292.6 ± 3.9 | 152.6 |

| | **Ni-33at.%Cr-0.1at%P** | | | | |
|---|---|---|---|---|---|
| As-received | 0 h | 144.6 ± 2.9 | 0.0 | | |
| 373 | 10 | 260.2 ± 6.7 | 115.6 | | |
| 373 | 126.5 | 271.6 ± 8.1 | 127.0 | | |
| 373 | 300 | 280.6 ± 9.0 | 136.0 | | |
| 373 | 1000 | 277.8 ± 5.2 | 133.2 | | |
| 373 | 3000 | 273.0 ± 9.0 | 128.4 | | |
| 418 | 3 | 168.0 ± 3.7 | 23.4 | | |
| 418 | 30 | 270.8 ± 3.3 | 126.2 | | |
| 418 | 100 | 280.4 ± 4.4 | 135.8 | | |
| 418 | 300 | 283.0 ± 6.0 | 138.4 | | |
| 418 | 1000 | 300.0 ± 5.3 | 155.4 | | |
| 418 | 3000 | 288.4 ± 11.3 | 143.8 | | |
| 470 | 1 | 255.4 ± 7.1 | 110.8 | | |
| 470 | 10 | 281.4 ± 7.1 | 136.8 | | |
| 470 | 30 | 285.2 ± 3.3 | 140.6 | | |
| 470 | 100 | 289.2 ± 6.4 | 144.6 | | |
| 470 | 300 | 286.6 ± 5.4 | 142.0 | | |
| 470 | 1000 | 295.6 ± 11.3 | 151.0 | | |
| 470 | 3000 | 299.6 ± 1.3 | 155.0 | | |

## S4. Kolmogorov–Johnson–Mehl–Avrami (KJMA) model fitting

The hardness data was also characterized by using the KJMA model (Eq. (9) in the manuscript). It was found that the data for all the alloys were well described by a common Avrami exponent, $n = 0.75$, and Arrhenius prefactor ($k_0 = 8.2\times10^7$) but by differing values for the peak hardness ($H_{max}$) and apparent activation energy, $Q$. As shown, increasing alloyed phosphorus increased the peak hardness and decreases the apparent activation energy. The fitting parameters are compared for each alloy in the table below.

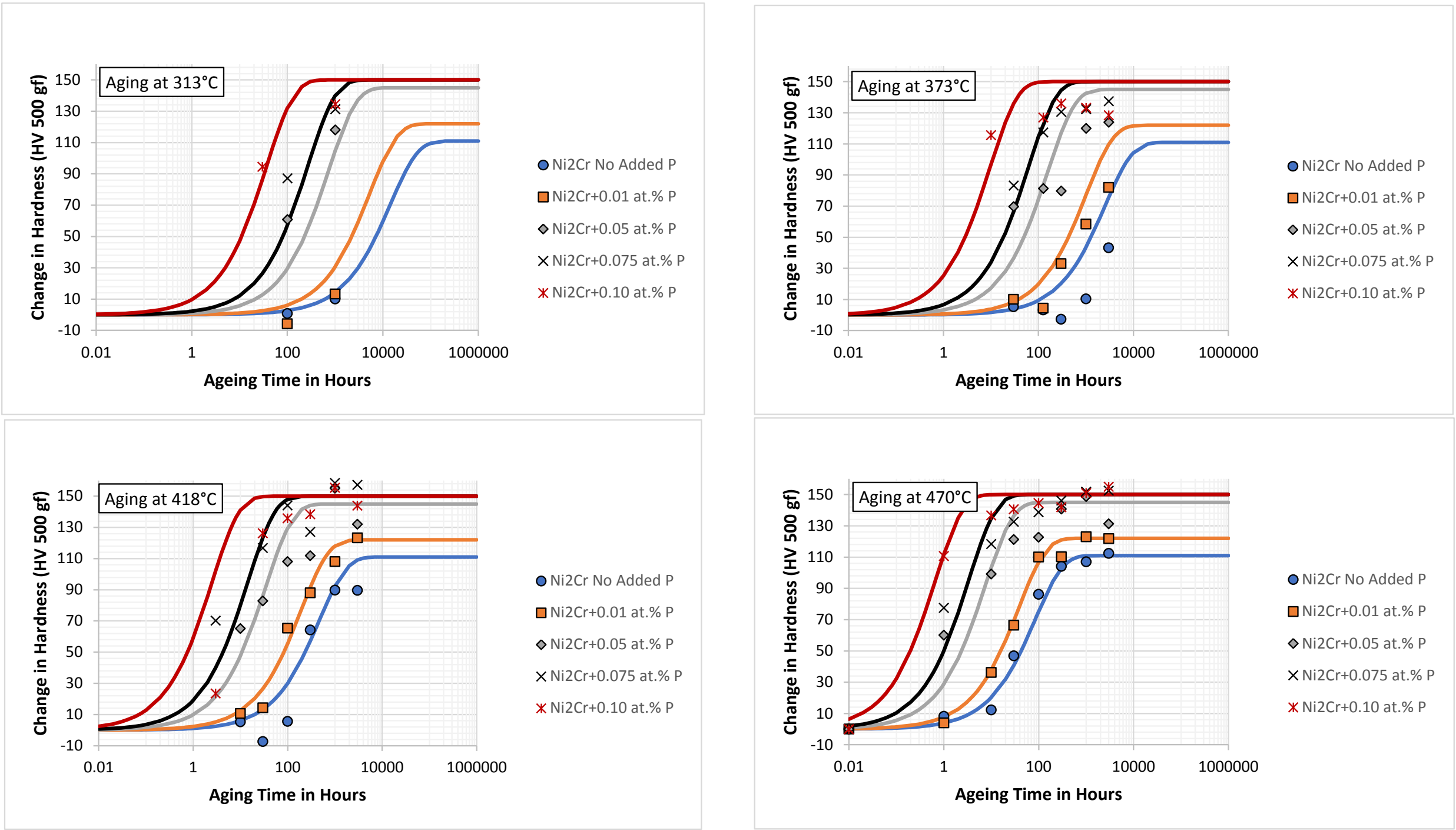

Figure S1. Summary of the isothermal aging experiments (data points) and trend lines based on the KJMA equation. Results show that alloyed phosphorus increases the hardening rate, the peak hardness but decreases the apparent activation energy for long range order.

Table S1. KJMA model parameters used to fit the micro-hardness data as shown in Figure S1.

| Alloy | $H_0$(HV) | $H_{max}$(HV) | n | $k_0$ | Q (eV) |
|---|---|---|---|---|---|
| Ni-33at.%Cr | 143 | 254 | 0.75 | $8.2x10^7$ | 1.45 |
| Ni-33at.%Cr + 0.01 at.% P | 143 | 265 | 0.75 | $8.2x10^7$ | 1.40 |
| Ni-33at.%Cr + 0.05 at.% P | 143 | 288 | 0.75 | $8.2x10^7$ | 1.29 |
| Ni-33at.%Cr + 0.075 at.% P | 143 | 293 | 0.75 | $8.2x10^7$ | 1.24 |
| Ni-33at.%Cr + 0.1 at.% P | 143 | 293 | 0.75 | $8.2x10^7$ | 1.14 |